\documentclass[longbibliography,groupedaddress,showpacs,showkeys,amssymb,eqsecnum,aps,nofootinbib]{revtex4}
\usepackage[pdftex]{graphicx}
\usepackage[pdftex]{graphics}
\usepackage{amsmath}
\usepackage{epsf}                       
\usepackage{color}                   
\usepackage{verbatim}                                                                         
\usepackage{hyperref}

\newcommand{\be}{\begin{equation}}

\newcommand{\ee}{\end{equation}}                                                                              
\newcommand{\fracd}[2]{\displaystyle\frac{#1}{#2}}

\begin{document}                                                                                              

\title{Structural identities in the first order formulation of quantum gravity}

\author{F. T. Brandt}  
\email{fbrandt@usp.br}
\affiliation{Instituto de F\'{\i}sica, Universidade de S\~ao Paulo, S\~ao Paulo, SP 05508-090, Brazil}

\author{J. Frenkel}
\email{jfrenkel@if.usp.br}
\affiliation{Instituto de F\'{\i}sica, Universidade de S\~ao Paulo, S\~ao Paulo, SP 05508-090, Brazil}

\author{S. Martins-Filho}   
\email{sergiomartinsfilho@usp.br}
\affiliation{Instituto de F\'{\i}sica, Universidade de S\~ao Paulo, S\~ao Paulo, SP 05508-090, Brazil}

\author{D. G. C. McKeon}
\email{dgmckeo2@uwo.ca}
\affiliation{
Department of Applied Mathematics, The University of Western Ontario, London, Ontario N6A 5B7, Canada}
\affiliation{Department of Mathematics and Computer Science, Algoma University,
Sault Ste.~Marie, Ontario P6A 2G4, Canada}

\date{\today}

\begin{abstract}
We study the self-consistency of the first order formulation of quantum
gravity, which may be attained by introducing, apart from the graviton field,
another auxiliary quantum field. By comparing the forms of the generating functional
$Z$ before and after integrating out the additional field, we derive a set of
structural identities which must be satisfied by the Green's functions at all
orders. These are distinct from the usual Ward identities, being necessary for
the self-consistency of the first order formalism. They relate the Green's
functions involving  the additional  quantum field to those containing a certain composite graviton field, which corresponds to its classical value. Thereby, the structural identities lead to a simple interpretation of the auxiliary field.
\end{abstract}                                                                                                

\pacs{11.15.-q}
\keywords{gauge theories; first order formulation; quantum gravity}

\maketitle


\section{Introduction}
The first order formulation of gauge theories has a simple form
involving only cubic interactions, which are
momentum-independent. This simplifies the computations of the quantum
corrections in the usual second-order gauge theories, that involve
momentum dependent three-point as well as higher-point vertices
\cite{Okubo:1979gt,Buchbinder:1985jcBuchbinder:1983ys,
McKeon:1994ds, Martellini:1997mu, Andrasi:2007dk, costello:2011b,  
Brandt:2015nxa, Brandt:2016eaj,anero2017oneloop,Frenkel:2017xvm, Frenkel:2018xup, Brandt:2018wxe}. In quantum gravity, for example, the first order formulation allows to replace an infinite number of complicated multiple graviton couplings present in the second-order Einstein-Hilbert (EH) action, by a small number of simple cubic vertices \cite{Brandt:2015nxa,Brandt:2016eaj}. The EH action has the form 
\begin{equation}\label{eq:11}
    S = - \frac{1}{16 \pi G_{N}} \int \mathop{d^d x} \sqrt{-g} g^{\mu \nu} R
    _{\mu \nu} (\Gamma),
\end{equation}
where $G_N$ is Newton's constant  and the affine connection $\Gamma_{\mu \nu}^{\lambda}$ may be written in terms of the metric $ g_{\mu \nu } $ as
\begin{equation}\label{eq:12}
    \Gamma_{\mu \nu}^{\lambda} = \frac{1}{2} g^{\lambda \sigma} \left (
    g_{ \mu \sigma, \nu } + g_{ \nu \sigma , \mu} - g_{ \mu \nu, \sigma} \right
    ). 
\end{equation}
The Ricci tensor $ R_{\mu \nu}( \Gamma ) $ is given by
\begin{equation}\label{eq:13}
    R_{\mu \nu} ( \Gamma ) = \Gamma_{ \mu \rho , \nu}^{ \rho} -
    \Gamma_{\mu \nu, \rho }^{\rho} - \Gamma_{ \mu \nu}^{ \sigma} \Gamma_{
    \sigma \rho}^{ \rho} + \Gamma_{ \mu \sigma}^{ \rho} \Gamma_{ \nu \rho}^{\sigma}.
\end{equation}
As noted by Einstein and Palatini 
\cite{Ferraris82}
at the classical level, it is possible to treat both $g_{
\mu \nu} $ and $ \Gamma_{ \mu \nu}^{ \lambda} $ as being independent
quantities. In this first order action, the equation of motion for $\Gamma_{\mu
\nu}^{ \lambda} $ yields Eq.~\eqref{eq:12}. At the quantum level, it has been shown \cite{Brandt:2015nxa,Brandt:2016eaj} that the radiative corrections computed using the first-order and second-order EH actions are the same.  

In a previous paper \cite{Brandt:2020sho}, we examined  a set of structural identities which are necessary for the consistency of the first order formulation of the Yang-Mills theory. The purpose of the present work is to extend this analysis to quantum gravity, where the corresponding structural identities ensure the self-consistency of the first order formulation.

To this end, we introduce a source $ j_{\mu \nu} $ for the graviton field and
also a source $ J_{ \mu \nu}^{ \lambda} $ for the other auxiliary field which
is treated as an independent  field,  and consider the generating functional
$Z[J,j] $ of  Green's functions. We then compare the functional dependence of
$Z[J,j]$ on the sources in the original first order formalism with that obtained after making a suitable shift which enables to integrate out the auxiliary field. The equality of these functional forms leads to a set of structural identities among the Green's functions which must be satisfied to all orders. Such identities are complementary but distinct from usual Ward identities, being necessary for the internal consistency of the first order formulation of quantum gravity.

These  identities show that in the first order formalism the Green's functions containing only external graviton fields are the same as the corresponding ones which occur in the second-order formulation. Furthermore, these identities relate the Green's functions involving external auxiliary fields to those involving a certain composite graviton field.
This combination, which corresponds to the classical value of the
auxiliary field, contains graviton fields which are pinched at the
same spacetime point. It is well known \cite{Wilson:1972ee,
  muta:book87, weinberg:book1995} that composite fields can lead to
short-distance singularities. In the present case, such singularities
are important for the cancellations of ultraviolet (UV) divergences arising from loop diagrams, which are necessary for the implementation of the structural identities.  

Since calculations in quantum gravity have a great algebraic
complexity, in section 2 we recast the analysis done in
\cite{Brandt:2020sho} into an alternative form, which is based on a
simpler diagonal representation of the first order formulation of the
Yang-Mills theory \cite{Brandt:2018avq}. 
Such a representation exhibits similar features to those in quantum
gravity, yet it is easier to handle algebraically. With this insight,
we consider in section 3 the Lagrangian and the generating functional
of Green's functions in a corresponding diagonal representation of the
first order formulation of quantum gravity \cite{Brandt:2016eaj}. 
In section 4 we derive a transparent structural identity, which has been explicitly verified to one-loop order, that clarifies the meaning of the auxiliary field in this formulation. In section 5, we study another structural identity satisfied by the Green's functions and examine the cancellations between the loop UV divergences and the short distance singularities arising  from the tree diagrams involving composite fields. A brief discussion of the results is given in section 6. Several details of the one-loop calculations are outlined in the Appendix.

\section{Structural identities in Yang-Mills theory}

The first order formulation of the Yang-Mills theory involves the gluon $
A_{\mu}^{a} $ and the auxiliary fields $F_{\mu \nu}^{a}$  whose dynamics is described by the Lagrangian           
\begin{equation}\label{eq:21}
    \tilde{\mathcal{L}}_{\textrm{YM}}^{(1)} = \frac{1}{4} F_{ \mu \nu}^{ a}
    F^{a \, \mu \nu} - \frac{1}{2} F_{}^{a \, \mu \nu} \left ( \partial_{\mu}
    A_{ \nu}^{ a} - \partial_{\nu} A_{ \mu}^{ a} + g f_{}^{ abc} A_{ \mu}^{ b} A_{ \nu}^{ c}  \right ).
\end{equation}
This form has a single vertex $ \left \langle FAA \right \rangle $ but leads to a rather involved non-diagonal matrix-propagator, containing the $(AA)$, $(FF)$ and the mixed $(FA)$, $(AF)$ propagators.
On the other hand, if we make in Eq.~\eqref{eq:21} the 
change of variable
\begin{equation}\label{eq:22}
    F_{ \mu \nu}^{ a} = \tilde{H}_{ \mu \nu}^{ a} + \partial_{\mu} A_{ \nu}^{ a} - \partial_{\nu} A_{ \mu}^{ a}
\end{equation}
one obtains the Lagrangian
\begin{equation}\label{eq:23}
    \tilde{\mathcal{L}}_{\textrm{YM}}^{\textrm{I}} =  \frac{1}{4}
    \tilde{H}_{\mu \nu}^{ a} \tilde{H}^{a \, \mu \nu} - \frac{1}{4}
    \left ( \partial_{\mu} A_{ \nu}^{ a} - \partial_{\nu} A_{ \mu}^{
            a} \right )^2 - \frac{g}{2} f_{}^{abc} \left ( \tilde{H}_{
            \mu \nu}^{ a} + \partial_{\mu} A_{ \nu}^{ a} - \partial_{\nu} A_{ \mu}^{
a} \right ) A_{}^{b \, \mu } A_{ }^{c \, \nu }, 
\end{equation}
which involves two cubic vertices $ \left \langle AAA \right \rangle $, $ \langle \tilde{H}AA \rangle $ as well as two simple propagators $(AA)$, $(\tilde{H} \tilde{H})$. The BRST renormalization of this diagonal formulation
of the Yang-Mills theory has been implemented to all orders in reference
\cite{Brandt:2018avq} (see also \cite{Buchbinder:2018jqs,Lavrov:2020exa,Barvinsky:2017zlx}). The complete Lagrangian density for this  formulation in covariant gauges is
\begin{equation}\label{eq:24}
    \mathcal{L}_{\textrm{YM}}^{ \textrm I} = \tilde{\mathcal{L}}_{\textrm{YM}}^{
    \textrm I} - \frac{1}{2 \xi}(\partial_{\mu} A^{\mu \, a} )^2 + \partial
    ^{\mu} \bar{\eta}^{a} (\delta^{ab} \partial_{\mu} - g f^{ abc}
    A^{c}_\mu ) \eta^{b}, 
\end{equation}
where $ \xi $ is a gauge-fixing parameter and $ \bar{\eta}^{a}$, $ \eta^{b} $ are ghost fields.     
In addition, we will also introduce the external sources $ \tilde{J}_{\mu \nu}^{a} $ and $ \tilde{j}^{a}_{\mu} $ as follows:
\begin{equation}\label{eq:25}
    \mathcal{L}_{\textrm{source}} = \tilde{J}_{ \mu \nu}^{ a} \tilde{H}_{}^{a \, \mu \nu} + \tilde{j}^{a}_{\mu} A^{a \, \mu}.
\end{equation}
The generating functional for Green's functions is given by the path integral
\begin{equation}\label{eq:26}
    Z[J,j] = N \int \mathcal{D} \eta \mathcal{D} \bar{\eta} \mathcal{D} \tilde{H} \mathcal{D} A \exp{i \left [S + \int \mathop{d^d x} \tilde{J}_{ \mu \nu}^{ a} \tilde{H}_{}^{a \, \mu \nu} + \tilde{j}^{a}_{\mu} A^{a \, \mu} \right ]}, 
\end{equation}
where $N$ is a normalization factor and $ S = \int \mathop{d^d x} \mathcal{L}^{ \textrm I}_{\textrm {YM}} $. This equation has a form which is suitable for functional differentiation with respect to $ \tilde{J}$ and $ \tilde{j} $, and therefore for obtaining the Green's functions.

If we were to set $ \tilde{J}^{a}_{\mu \nu} = 0 $ at the outset (so that we would consider Green's functions with only external fields $A^{a}_{\mu}$) and make the change of variable in the functional integral
\begin{equation}\label{eq:27}
    \tilde{H}^{a}_{\mu \nu} \rightarrow \tilde{H}^{a}_{\mu \nu} + g f^{abc} A^{b}_{\mu} A^{c}_{\nu}
\end{equation}
then one can integrate out the $ \tilde{H}^{a}_{\mu \nu} $ field and find that
\begin{equation}\label{eq:28}
    Z[\tilde{J} = 0 , \tilde{j}] = Z_2[ \tilde{j}],
\end{equation}
where $Z_{2}[ \tilde{j}]$ is the generating functional for the second-order theory, characterized by the Lagrangian density
\begin{equation}\label{eq:29}
    \mathcal{L}^{\textrm {II}}_{\textrm{YM}} = - \frac{1}{4} ( \partial_{\mu}
    A_{ \nu}^{ a} - \partial_{\nu} A_{ \mu}^{ a} + g f_{}^{ abc} A_{
      \mu}^{ b} A_{ \nu}^{ c}  )^2 - \frac{1}{2 \xi}(\partial_{\mu}
    A^{\mu \, a} )^2 
+ \partial^{\mu} \bar{\eta}^{a} (\delta^{ab} \partial_{\mu} 
- g f^{ abc}A^{c}_\mu ) \eta^{b}
\end{equation}
together with the source term $ \tilde{j}_{\mu}^{a} A^{a \, \mu}$. This establishes the important property that the Green's functions with only external gluon fields are the same in both approaches.

We now consider using $ Z[\tilde{J}, \tilde{j}]$ with $ \tilde{J} \neq 0 $ and examine what changes occur in the first order formalism when there are external fields $ \tilde{H}_{ \mu \nu}^{ a}$. To this end, we consider
in place of Eq.~\eqref{eq:27}, the shift  
\begin{equation}\label{eq:210}
    \tilde{H}^{a}_{\mu \nu} \rightarrow \tilde{H}^{a}_{\mu \nu} + g f^{abc} A^{b}_{\mu} A^{c}_{\nu} - 2 \tilde{J}^{a}_{\mu \nu}.
\end{equation}
This leads, after integrating out the $ \tilde{H}^{a}_{\mu \nu} $ field, to the alternative form of the generating functional
\begin{equation}\label{eq:211}
    Z'[J,j] = N \int \mathcal{D} \eta \mathcal{D} \bar{\eta} \mathcal{D} A \exp{i \left [ \int \mathop{d^d x} \left ( \mathcal{L}^{\textrm{II}}_{\textrm{YM}} + g f^{abc} \tilde{J}_{ \mu \nu}^{ a} A^{b \, \mu } A^{c \, \nu} - \tilde{J}_{ \mu \nu}^{ a} \tilde{J}_{}^{ a \, \mu \nu }  + \tilde{j}^{a}_{\mu} A^{a \, \mu} \right )\right ]}.
\end{equation}
This equals to $Z_{2}[ \tilde{j} ]$ in Eq.~\eqref{eq:28} if we set $ \tilde{J}_{\mu \nu}^{a} = 0$. It is interesting to note the unusual dependence of $Z'[ \tilde{J}, \tilde{j}]$ on $ \tilde{J} $ in Eq.~\eqref{eq:211}. 

Comparing the forms Eq.~\eqref{eq:26} and Eq.~\eqref{eq:211} of the   generating functionals and differentiating these with respect to $ \tilde{J}$ and $ \tilde{j}$, leads to a set of structural identities among the
Green's functions, which must be satisfied to all orders. Such structural identities lead to relations between the Green's functions involving $ \tilde{H}$ fields and the Green's functions that contain the composite fields $g f^{abc} A^{b}_{\mu}(x)A^{c}_{\nu}(x)$.
These identities hold both for the finite as well as for the UV divergent parts of the Green's functions. We have verified them explicitly  for the divergent contributions to one-loop order, using dimensional regularization in $4-2 \epsilon $ dimensions.

Taking the functional derivatives of Eq.~\eqref{eq:26} and Eq.~\eqref{eq:211} with respect to $ \tilde{J}^{a\,  \mu \nu} (x)$ and $ \tilde{j}^{b\,  \alpha}  (y)$ at $ \tilde{J}= \tilde{j}= 0$ and equating the results, we obtain the relation
\begin{equation}\label{eq:212}
    \langle 0 | T \tilde{H}^{a}_{\mu \nu} (x) A_{\alpha}^{b} (y) |0 \rangle = g f^{ade} \langle 0 | T A^{d}_{\mu} (x) A_{\nu}^{e} (x) A_{\alpha}^{b} (y) |0 \rangle. 
\end{equation}
Eq.~\eqref{eq:212} represents a quantum mechanical extension of the relation: $ \tilde{H}^{a}_{\mu \nu} = g f^{abc} A^{b}_{\mu} A^{c}_{\nu}$ which holds at the classical level. This structural identity is clearly satisfied in the tree
approximation, since the mixed $( \tilde{ H} A)$ propagator vanishes in our theory. The right hand side of Eq.~\eqref{eq:212} also vanishes at the tree level. To order $g^2$, we find that the divergent parts on both sides of Eq.~\eqref{eq:212} are, in momentum space, equal to
\begin{equation}\label{eq:213}
    - \frac{g^{2} C_{\textrm{YM}}}{16 \pi^{2} \epsilon} \frac{5 + \xi}{4} \frac{\delta^{ab}}{k^2} 
    \left ( k_{\mu} \eta_{\nu \alpha} - k_{\nu} \eta_{\mu \alpha} \right ).
\end{equation}
Applying $ \delta^{2} / \delta \tilde{J}^{a\, \mu \nu} (x) \delta \tilde{J}^{b\, \alpha \beta} (y) $ to Eqs.~\eqref{eq:26} and \eqref{eq:211} and equating the results, leads to
\begin{equation}\label{eq:214}
    \langle 0| T \tilde{H}_{ \mu \nu}^{ a}(x) \tilde{H}_{ \alpha \beta}^{ b}
    (y)| 0\rangle = 2i I_{\mu \nu , \alpha \beta} \delta^d(x-y) + g^{2} f^{ab'c'} f^{bd'e'} \langle 0|T A^{b'}_{\mu} (x) A^{c'}_{\nu} (x) A^{d'}_{\alpha} (y) A^{e'}_{\beta} (y) | 0 \rangle,
\end{equation}
where
\begin{equation}\label{eq:215}
    I_{\mu \nu , \alpha \beta} = \frac{1}{2} \left ( \eta_{\mu \alpha} \eta_{\nu \beta} - \eta_{\nu \alpha} \eta_{\mu \beta} \right ).
\end{equation}
This identity is also manifestly satisfied at three level, where the first term on the right hand side of Eq.~\eqref{eq:214} is just equal to the tree propagator $(\tilde{H}\tilde{H})$. To order $g^{2} $, one can verify, in momentum space, that the divergent part on both sides of Eq.~\eqref{eq:214} are equal to
\begin{equation}\label{eq:216}
    - i\frac{g^{2} C_{\textrm{YM}}}{16 \pi^{2} \epsilon} (1 + \xi) \delta^{ab} I_{\mu \nu, \alpha \beta}.
\end{equation}

It is worth to point out that in the identities Eq.~\eqref{eq:212} and Eq.~\eqref{eq:214}, the origin of the divergent contributions is different. On the left hand side of these equations, UV divergences come from one-loop graphs, whereas on their right hand side short-distance singularities arise from the pinched tree graphs.

Further differentiations of Eqs.~\eqref{eq:26} and \eqref{eq:211} with respect to $ \tilde{J}$ and $ \tilde{j} $ yield a set of structural identities which are complementary to the usual Ward identities.
One can compare the above identities with the ones found in 
reference \cite{Brandt:2020sho} in the usual first order formulation of the Yang-Mills theory. (see, for example, equations (3.1); (3.2) and (4.1); (4.2) in \cite{Brandt:2020sho}). One can see that the structural identities obtained in the  diagonal representation have a much simpler form. This feature will be especially useful for the derivation of the corresponding identities in quantum gravity.

\section{Diagonal formulation of first order Palatini action}

Instead of using $ g_{ \mu \nu} $ and $ \Gamma_{\mu \nu}^{\lambda}$ as independent fields in the action Eq.~\eqref{eq:11}, it turns out
to be more useful to employ the independent combinations \cite{Brandt:2015nxa}
\begin{equation}\label{eq:31}
    h^{\mu \nu} = \sqrt{-g} g^{\mu \nu}
\end{equation}
and  
\begin{equation}\label{eq:32}
    G_{ \mu \nu}^{ \lambda} = \Gamma_{ \mu \nu}^{ \lambda} - \frac{1}{2} \left ( \delta^{\lambda}_{\mu} \Gamma_{ \nu \sigma}^{ \sigma} + \delta_{\nu}^{\lambda} \Gamma_{ \mu \sigma}^{ \sigma} \right ).
\end{equation}
Thus, we arrive at the following Lagrangian density
in $d$ spacetime dimensions
\begin{equation}\label{eq:33}
    \tilde{\mathcal{L}}^{(1)}_{\textrm{EH}} = \frac{h^{\mu \nu}}{\kappa^2} \left (
    {G}_{ \mu \nu, \lambda}^{ \lambda} + \frac{1}{d-1} {G}_{ \mu \lambda}^{
\lambda} {G}_{ \nu \sigma}^{ \sigma} - {G}_{ \mu \sigma}^{ \lambda}
{G}_{ \nu \lambda}^{ \sigma} \right ).
\end{equation}
In order to proceed, $h^{\mu \nu}$ is expanded about a flat metric $ \eta^{\mu \nu} $ ($ \kappa  = \sqrt{16 \pi G_N} $) 
\begin{equation}\label{eq:36}
    h^{\mu \nu} (x) = \eta^{\mu \nu} + \kappa \phi^{\mu \nu} (x). 
\end{equation}

Eq.~\eqref{eq:33} yields a basic vertex $ \left \langle
\phi GG \right \rangle $ (see Eqs. \eqref{eq:36} and \eqref{eq:39} below). 
However, it leads to an involved non-diagonal matrix
propagator containing $(\phi \phi)$, $(GG)$ 
and the mixed propagator
$(\phi G)$. 
As in the Yang-Mills theory, it proves convenient to use a diagonal formulation
of the first order EH action \cite{Brandt:2016eaj}. This may be achieved by
making the change of variable (compare with Eq.~\eqref{eq:22})
\begin{equation}\label{eq:34}
    {G}_{ \mu \nu}^{ \lambda} = {H}_{ \mu \nu}^{ \lambda} + (M^{-1})_{\mu
    \nu}^{\lambda} {}_{ \pi \tau}^{ \rho} (h = \eta ) h^{\pi \tau}_{,
\rho},
\end{equation}
where
\begin{equation}\label{eq:35}
    (M^{-1})_{\mu \nu}^{\lambda} {}_{ \pi \tau}^{ \rho} (h) = -
    \frac{1}{2(d-2)} h^{\lambda \rho} h_{\mu \nu} h_{\pi \tau} +
    \frac{1}{4} h^{\lambda \rho} \left ( h_{\pi \mu} h_{\tau \nu} + h_{\pi \nu} h_{\tau \mu}\right ) - \frac{1}{4} \left ( h_{\tau \mu} \delta_{\nu}^{\rho} \delta_{\pi}^{\lambda} + h_{\pi \mu} \delta_{\nu}^{\rho} \delta_{\tau}^{\lambda} +  h_{\tau \nu} \delta_{\mu}^{\rho} \delta_{\pi}^{\lambda} +  h_{\pi \nu} \delta_{\mu}^{\rho} \delta_{\tau}^{\lambda}\right ). 
\end{equation}
In this way, the Lagrangian density Eq.~\eqref{eq:33} may be written in the form
\begin{equation}\label{eq:37}
\begin{split}
    \tilde{\mathcal{L}}^{ \textrm{I}}_{ \textrm{EH}}& = \frac{1}{2}
    {H}_{ \mu \nu}^{ \lambda} M_{\lambda}^{\mu \nu} {}_{\rho}^{ \pi
      \tau}(\eta)
 {H}_{ \pi \tau }^{ \rho} - \frac 1 2 {\phi}_{ , \lambda}^{ \mu \nu} (M^{-1})_{\mu \nu}^{\lambda} {}_{ \pi \tau}^{ \rho} (\eta ) {\phi}_{ , \rho}^{ \pi \tau} \\ &+ \frac{\kappa }{2} [ {H}_{ \mu \nu}^{ \lambda} + {\phi}_{ , \rho}^{ \alpha \beta} (M^{-1})_{\alpha \beta}^{\rho} {}_{ \mu \nu}^{ \lambda}(\eta)] M^{\mu \nu}_{\lambda} {}_{ \sigma }^{ \pi \tau} (\phi ) \left [ {H}_{\pi \tau }^{\sigma } + (M^{-1})_{\pi \tau }^{\sigma } {}_{ \gamma \delta }^{ \upsilon }(\eta){\phi}_{ , \upsilon }^{ \gamma \delta}  \right ].
 \end{split}
\end{equation}
where $M^{\mu \nu}_{\lambda} {}_{\sigma }^{ \pi \tau} (\phi )$ 
is given by
\begin{eqnarray}\label{eq:36BM2016}
    M^{\mu\nu}_{\lambda}{}^{\pi\tau}_{\sigma}(\phi)   & = &
\frac{1}{2}\left[\frac{1}{d-1}\left( \delta^\nu_\lambda\delta^\tau_\sigma \phi^{\mu\pi}+
                                                \delta^\mu_\lambda\delta^\tau_\sigma \phi^{\nu\pi}+
                                                \delta^\nu_\lambda\delta^\pi_\sigma \phi^{\mu\tau}+
                                                \delta^\mu_\lambda\delta^\pi_\sigma \phi^{\nu\tau}
\right) 
 \right.  \nonumber \\  && - \left. 
\left( 
                                                \delta^\tau_\lambda\delta^\nu_\sigma \phi^{\mu\pi}+
                                                \delta^\tau_\lambda\delta^\mu_\sigma \phi^{\nu\pi}+
                                                \delta^\pi_\lambda\delta^\nu_\sigma \phi^{\mu\tau}+
                                                \delta^\pi_\lambda\delta^\mu_\sigma \phi^{\nu\tau}
\right) \frac{}{} \!\!\right]. 
\end{eqnarray}
Thus, we see that the Lagrangian Eq.~\eqref{eq:37} involves three
cubic vertices $ \langle H \phi H\rangle$, $ \langle \phi H
\phi\rangle$  and $ \langle \phi \phi \phi\rangle $. On the other
hand,  it leads only to two uncoupled propagators $ (\phi \phi )$ and
$ (HH) $. 

Using the Lagrangian in Eq. \eqref{eq:37} in the 
Euler-Lagrange equation for the field $H^\lambda_{\mu\nu}$
we obtain the classical solution
\be
 {H}_{ \mu \nu}^{ \lambda} = -\left[\left(
 M(\eta) + \kappa M(\phi) \right)^{-1} \kappa M(\phi) M^{-1}(\eta)
\right]{}^\lambda_{\mu\nu}  {}_{ \pi \tau}^{ \rho} \phi_{, \rho}^{\pi \tau}.
\ee
Since 
$M(\eta) + \kappa M(\phi) = M(\eta+\kappa\phi)$,
this can be written as
\begin{equation}\label{eq:38}
   {H}_{ \mu \nu}^{ \lambda} = [M^{-1}(\eta + \kappa \phi ) - M^{-1}(\eta ) ] _{\mu \nu}^{\lambda} {}_{ \pi \tau}^{ \rho} \phi_{, \rho}^{\pi \tau}.
\end{equation}
Substituting \eqref{eq:38} back into \eqref{eq:37}, we obtain 
(using $\phi^{\mu\nu}_{,\lambda} = h^{\mu\nu}_{,\lambda}$)
\be
-\frac 1 2 h^{\mu\nu}_{,\lambda}
(M^{-1})_{\mu\nu}^\lambda {}^\rho_{\pi\tau}(h) h^{\pi\tau}_{,\rho}
\ee
which is just the classical second-order Einstein-Hilbert Lagrangian. 
This demonstrate the
classical equivalence of the two formalisms \cite{Brandt:2016eaj}.

In order to obtain the propagator of the  $\phi^{\mu\nu}$ field, we use the gauge fixing Lagrangian
\begin{equation}\label{eq:39}
{\cal L}_{\textrm{gf}} = -\frac{1}{2\xi}(\phi^{\mu\nu}_{,\nu})^2.
\end{equation}
With this gauge fixing,  the contributions coming from 
the vector ghost fields     $d_\nu$, $\bar d_\mu$           are \cite{Capper:1973pv}

\begin{eqnarray}\label{eq:310}
    {\cal L}_{\textrm{ghost}} &=& 
\bar d_\mu\left[\partial^2 \eta^{\mu\nu}
+(\phi^{\rho\sigma}_{,\rho})\partial_\sigma\eta^{\mu\nu}
-(\phi^{\rho\mu}_{,\rho})\partial^{\nu} 
+\phi^{\rho\sigma}\partial_{\rho}\partial_\sigma\eta^{\mu\nu}
-(\partial_\rho\partial^\nu\phi^{\rho\mu})\right] d_\nu
\end{eqnarray}
Thus, the complete diagonal first order 
Lagrangian density becomes
\begin{equation}\label{eq:311}
    \mathcal{L}^{\textrm{I}}_{\textrm{EH}} =
    \tilde{\mathcal{L}}_{\textrm{EH}}^{\textrm{I}} 
+ {\cal L}_{\textrm{gf}} 
+ \mathcal{L}_{\textrm{ghost}}.
\end{equation}

Next, we will also introduce the external sources              
$J^{\mu\nu}_\lambda$ and $j_{\mu\nu}$
as follows
\begin{equation}\label{eq:312}
    \mathcal{L}_{\textrm{source}} = {J}_{\lambda}^{\mu \nu} {H}_{\mu \nu}^{\lambda} + {j}_{\mu \nu}^{} \phi^{\mu \nu}.
\end{equation}
Using the above results, the generating functional 
for Green's functions will
be given by the Feynman path integral
\begin{equation}\label{eq:313}
    Z[J,j] = N \int \mathop{\mathcal{D} d} \mathop{\mathcal{D}
      \bar{d}} 
\mathop{\mathcal{D} H}
{\mathop \mathcal{D} \phi} \exp{i \left [S + \int \mathop{d^d x} \left ({J}_{\lambda}^{\mu \nu} {H}_{\mu \nu}^{\lambda} + {j}_{\mu \nu}^{} \phi^{\mu \nu} \right)\right ]}, 
\end{equation}
where $N$ is a normalization factor and 
$S=\int d^d x {\cal L}^\textrm{I}_{\textrm{EH}}$.           
This equation has a form which is appropriate for generating 
the Green's functions,  through the application 
of functional differentiations with respect to
$J^{\mu\nu}_\lambda$ and $j_{\mu\nu}$.

Performing the following shift in the functional integral \eqref{eq:313}
\begin{equation}\label{eq:314}
    {H}_{ \mu \nu}^{ \lambda} \rightarrow {H}_{ \mu \nu}^{ \lambda} +
    [M^{-1}(\eta + \kappa \phi ) - M^{-1}(\eta ) ] _{\mu
      \nu}^{\lambda} {}_{ \pi \tau}^{ \rho} \phi^{\pi \tau}_{,\rho} -
    (M^{-1})_{\mu\nu}^{\lambda} {}_{ \pi \tau}^{ \rho} (\eta + \kappa
    \phi ) {J}_{\rho}^{\pi \tau} ,
\end{equation}
we obtain
\begin{equation}\label{eq:315a}
    \begin{split} 
        Z'[J,j] = N \int \mathop{\mathcal{D} d} \mathop{\mathcal{D} \bar{d}}
        {\mathop \mathcal{D} \phi} \mathop{\mathcal{D} H} 
\exp i \int \mathop{d^d x} \{
\frac 1 2 H_{\mu\nu}^\lambda [M(\eta) +\kappa M(\phi)]^{\mu\nu}_\lambda  {}^{\pi\tau}_\sigma H_{\pi\tau}^\sigma +
        \mathcal{L}_{\textrm{EH}}^{\textrm{II}} \\+  {J}_{\lambda}^{\mu \nu}
[M^{-1}(\eta + \kappa \phi ) - M^{-1}(\eta ) ] _{\mu \nu}^{\lambda} {}_{ \pi \tau}^{ \rho} 
\phi^{\pi\tau}_{,\rho}  - \frac{1}{2} {J}_{\lambda}^{\mu \nu} (M^{-1})_{\mu\nu}^{\lambda} {}_{ \pi \tau}^{ \rho} (\eta + \kappa \phi ) {J}_{\rho}^{\pi \tau}+ {j}_{\mu \nu}^{} \phi^{\mu \nu} \}, 
    \end{split} . 
\end{equation}
This enables to integrate out the auxiliary 
field $H^\lambda_{\mu\nu}$ and leads to the alternative form 
of the generating functional \footnote{The Green's functions
                             which involves the field
                             $H^\lambda_{\mu\nu}$
                             necessarily will have
                             $H^\lambda_{\mu\nu}$ appearing in a
                             closed loop. But the propagator for the
                             field $H^\lambda_{\mu\nu}$ is momentum
                             independent and hence the associated loop momentum
                             integrals vanish if we
                             use dimensional regularization \cite{Brandt:2016eaj}. }
\begin{equation}\label{eq:315}
    \begin{split} 
        Z'[J,j] = N \int \mathop{\mathcal{D} d} \mathop{\mathcal{D} \bar{d}}
        {\mathop \mathcal{D} \phi}  
\exp i \int \mathop{d^d x} \{
        \mathcal{L}_{\textrm{EH}}^{\textrm{II}} +  {J}_{\lambda}^{\mu \nu}
[M^{-1}(\eta + \kappa \phi )  - M^{-1}(\eta ) ] _{\mu \nu}^{\lambda} {}_{ \pi \tau}^{ \rho} 
\phi^{\pi\tau}_{,\rho}  \\- \frac{1}{2} {J}_{\lambda}^{\mu \nu} (M^{-1})_{\mu\nu}^{\lambda} {}_{ \pi \tau}^{ \rho} (\eta + \kappa \phi ) {J}_{\rho}^{\pi \tau}+ {j}_{\mu \nu}^{} \phi^{\mu \nu} \}, 
    \end{split} . 
\end{equation}
where ${\cal L}^{\textrm{II}}_{\textrm{EH}}$
is the second-order EH Lagrangian, with ghosts and gauge fixing,
which may be written as
\begin{equation}\label{eq:316}
    \mathcal{L}_{\textrm{EH}}^{\textrm{II}} = - \frac{1}{2} {\phi}_{, \lambda}^{\mu \nu} (M^{-1})_{\mu\nu}^{\lambda} {}_{ \pi \tau}^{ \rho} (\eta + \kappa \phi ){\phi}_{, \rho}^{\pi \tau} - \frac{1}{2 \xi} ( {\phi}_{, \nu}^{\mu \nu})^{2} + \mathcal{L}_{\textrm{ghost}}. 
\end{equation}
We remark that the alternative generating 
functional \eqref{eq:315} has a certain similarity 
to the corresponding functional in the Yang-Mills theory given by
\eqref{eq:211}. The analogy is even
 more pronounced if we note that the coefficient of the source $J$ is
 just the result 
found at the classical level, given in Eq. \eqref{eq:38}
for the auxiliary field. We also note though that unlike Eq. \eqref{eq:211},
the term quadratic in the source $J$ for the auxiliary field 
contains field dependency. 
Using a similar procedure to that employed in the Yang-Mills theory [see Eqs.~\eqref{eq:27}--\eqref{eq:29}], one can show that the Green's function with only external gravitons are the same in the first and second-order formulations.

\section{Consistency condition for the auxiliary field}
Taking the functional derivatives of Eqs. \eqref{eq:313} and \eqref{eq:315} with respect to $J^{\mu\nu}_\lambda$ and $j_{\pi\tau}$ at $J^{\mu\nu}_\lambda = j_{\pi\tau} = 0$ and equating the results, we obtain the structural identity 
\begin{equation}\label{eq:41}
    \langle 0|T {H}_{\mu \nu}^{\lambda} ( x) {\phi}^{\pi \tau} (
    y) | 0 \rangle = \langle 0|T
[M^{-1}(\eta + \kappa \phi ) - M^{-1}(\eta ) ] _{\mu \nu}^{\lambda} {}_{ \alpha \beta}^{ \rho} 
     \phi^{\alpha\beta}_{,\rho}(x){\phi}^{\pi \tau} ( y)| 0 \rangle,
\end{equation}
where $(M^{-1})^{\lambda}_{\mu\nu}{}^\rho_{\alpha\beta}$
is defined in Eq. \eqref{eq:35}.
Equation \eqref{eq:41} is manifestly satisfied 
at tree level because its left hand side vanishes since 
there is no mixed $H\phi$ propagator in the theory. 
Similarly, the right hand side of \eqref{eq:41} 
vanishes in the tree approximation (order zero in $\kappa$).

\begin{figure}[t]
\includegraphics[scale=0.8]{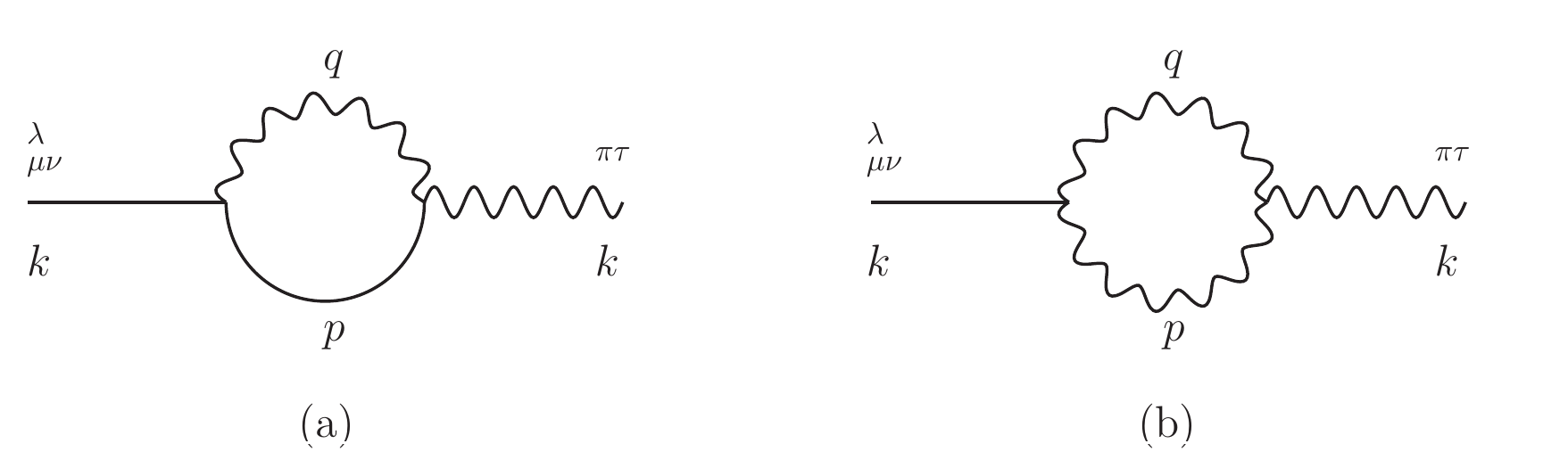} 
\caption{One-loop contributions to the propagator 
($H^\lambda_{\mu\nu} \phi^{\pi\tau}$). We are free to choose either $p=q-k$ 
or $q$ as the loop integration momentum.}
\label{fig1}
\end{figure}

To one-loop order,  the contribution to the left hand 
side  of Eq. \eqref{eq:41} arises from the 
Feynman diagrams shown in Fig. \ref{fig1}.
Using dimensional regularization, the contribution 
from the graph in Fig. (1a) actually vanishes, 
while the divergent contribution coming from 
graph in Fig. (1b) is given in momentum space,  
in the gauge $\xi=1$ by 
\be\label{eq:42}
\frac{- \kappa^2}{16\pi^2\epsilon}\left[
\frac{31}{96} k^\lambda\left(
\delta^{\pi}_{\mu} \delta^{\tau}_{\nu} +
\delta^{\pi}_{\nu} \delta^{\tau}_ {\mu} 
\right) + \dots
\right] ,
\ee
where $\dots$ stands for terms with other tensor structures which are
given in a general gauge in the Appendix.



\begin{figure}[t]
\includegraphics[scale=0.8]{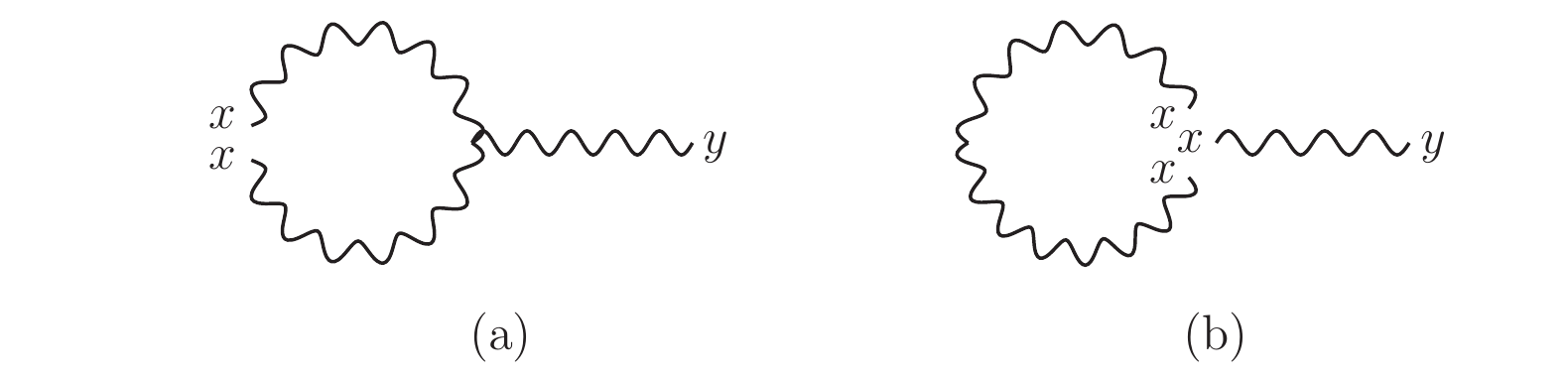} 
\caption{Pinched contributions to the right hand side of Eq. \eqref{eq:41}.}
\label{fig2}
\end{figure}

In order to obtain the contribution coming from the right hand side of
Eq. \eqref{eq:41}, one must expand the expression 
in the square bracket in a power 
series of $\kappa \phi$.   
Using for simplicity a schematic notation, we obtain
\be\label{eq:43}
M^{-1}(\eta+\kappa\phi)-M^{-1}(\eta)=
-\kappa M^{-1}(\eta) M(\phi) M^{-1}(\eta)  
+\kappa^2 M^{-1}(\eta) M(\phi) M^{-1}(\eta) M(\phi)  M^{-1}(\eta)  
+ \cdots,
\ee
where $M(\phi)$ is  a linear function of $\phi$
which is given by Eq. \eqref{eq:36BM2016}.
Substituting this result in the right hand side of 
Eq. \eqref{eq:41} one gets,  up to order $\kappa^2$,
two terms that involve, respectively, a product of 
three and four $\phi$  fields.
Using Wick's theorem
we can verify that the contribution from the cubic term comes from the  Feynman graph shown in Fig. \ref{fig2}a.
This diagram corresponds to a three-point tree Green's function
which has however two coordinates pinched 
at the same space time point $x$.


As we have mentioned earlier, such a composite 
field leads to an ultra-violet (short distance) contribution.
Using the appropriate expression for the three-point graviton
vertex \cite{Brandt:2016eaj},
one can evaluate in momentum space 
the contribution from Fig. (2a). 
The result turns out to 
be in agreement 
with the one given in 
Eq. \eqref{eq:42}.  
One must also consider the contribution involving 
four      $\phi$
fields in \eqref{eq:41}, 
which arises due to the last term in Eq.
\eqref{eq:43}.
This is represented by the Feynman diagram shown in 
figure \ref{fig2}b.
However, such a pinched contribution vanishes upon using 
dimensional regularization.

Thus, we see that the features which appear in the structural 
identity \eqref{eq:41} are similar to 
those which occur in the Yang-Mills theory via the 
identity \eqref{eq:212}. It is straightforward to generalize 
Eq. \eqref{eq:41} to an arbitrary number of graviton fields, 
namely
\begin{equation}\label{eq:44}
     \langle 0| T H_{ \mu \nu}^{ \lambda} (x) \phi^{\pi_{1}
     \tau_{1}}(y_{1} ) \cdots \phi^{\pi_{n} \tau_{n}} (y_{n} )| 0
     \rangle = \langle 0| T
[M^{-1}(\eta + \kappa \phi ) - M^{-1}(\eta ) ] _{\mu \nu}^{\lambda} {}_{ \alpha \beta}^{ \rho}  \phi_{, \rho}^{\alpha \beta} (x) \phi^{\pi_{1}
 \tau_{1}}(y_{1} ) \cdots \phi^{\pi_{n} \tau_{n}} (y_{n} ) | 0
         \rangle.
 \end{equation}
This relation may be interpreted as being, in quantum gravity, a
quantum-mechanical extension  of the relation \eqref{eq:38} which holds at the classical level.

\section{A second structural identity}
\begin{figure}[t]
\includegraphics[scale=0.8]{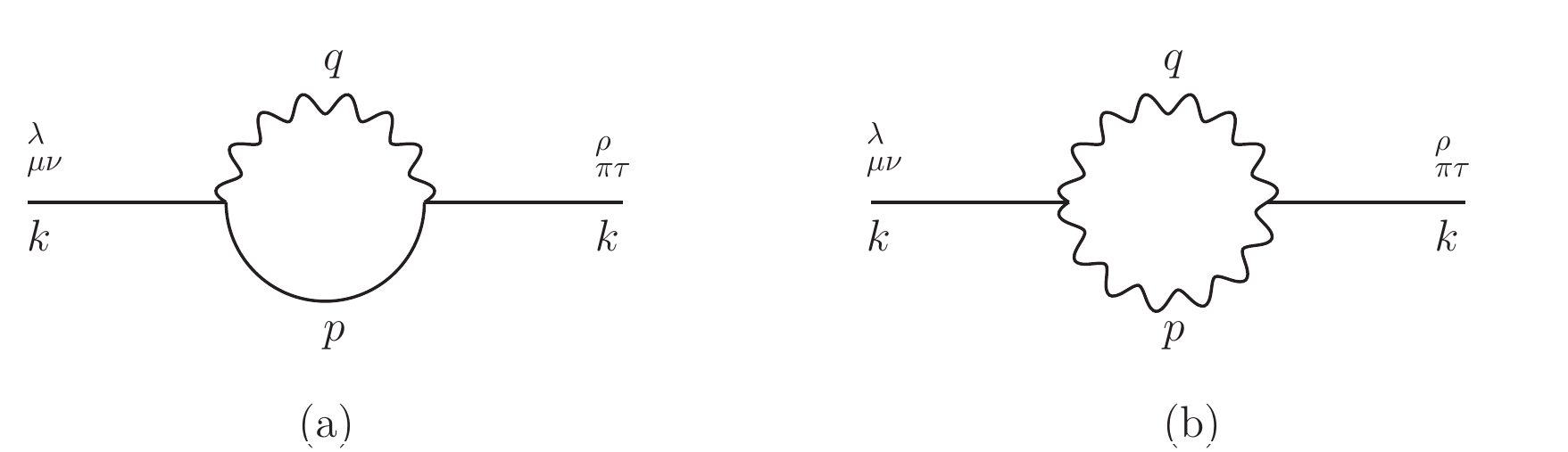} 
\caption{One-loop contributions to the propagator 
($H^\lambda_{\mu\nu} H^\rho_{\pi\tau}$).}
\label{fig3}
\end{figure}

Applying                                    
$
\delta^2/\delta J^{\mu\nu}_\lambda(x) \delta J^{\pi\tau}_\rho(y)
$
to Eqs. \eqref{eq:313} and \eqref{eq:315} 
and equating the results, yields
\begin{equation}\label{eq:51}
    \langle 0|T {H}_{\mu \nu}^{\lambda} ( x) {H}_{\pi \tau}^{\rho } (
    y)  | 0 \rangle = i \langle 0|T ( M^{-1})_{\mu \nu}^{\lambda}
    {}_{\pi \tau}^{\rho} (\eta + \kappa \phi )(x) | 0 \rangle\, 
\delta^{d} (x - y) 
+ \langle 0|T {\Delta}_{\mu \nu}^{ \lambda } [\phi (x)] {\Delta}_{\pi \tau}^{\rho} [ \phi (y) ] | 0 \rangle,
\end{equation}
where we have introduced the shorthand notation
\begin{equation}\label{eq:52}
    \Delta_{\mu \nu}^{\lambda} [\phi (x)] = 
    [M^{-1}(\eta + \kappa \phi ) - M^{-1}(\eta ) ] _{\mu \nu}^{\lambda} {}_{ \pi \tau}^{ \rho}  \phi_{, \rho}^{\pi \tau} (x).
\end{equation}
The structural identity \eqref{eq:51} 
is clearly satisfied at tree level, where the ($H H$) 
propagator is precisely 
equal to
$(M^{-1})^\lambda_{\mu\nu}{}^\rho_{\pi \tau}(\eta)\delta^d(x-y)$ 
We will now examine the perturbative expansion 
of each side of Eq. \eqref{eq:51}.
To one-loop order, the contributions to 
the left hand side of this equation arise from 
the Feynman diagrams shown in Fig. \ref{fig3}.

Using dimensional regularization, the contribution 
from graph in Fig. (\ref{fig3}a) vanishes, while the divergent 
part of the contribution from graph in Fig. (\ref{fig3}b) 
is given in momentum space,
in the gauge $\xi=1$, by
\be\label{eq:53}
\frac{i \kappa^2 k^2}{16\pi^2\epsilon}\left[
\frac{1}{24}\delta^\lambda_\nu\delta^\rho_\tau
\left(
\eta_{\pi\mu} + 6\frac{k_\pi k_\mu}{k^2}
\right) + \dots
\right],
\ee
where $\dots$ denote terms with other tensorial structures which are
explicitly given in the Appendix.

Next, let us examine the contributions of 
order $\kappa^2$ which come from the terms 
on the right hand side of the Eq. \eqref{eq:51}.
Such a contribution could arise from the first term, but
this vanishes upon using dimensional regularization.
Thus, we must evaluate only the
$\kappa^2$
contribution coming from the last term.
This part arises by considering the terms
of order $\kappa$ which occur in each of the factors 
appearing in  the last expression on the
right hand side of the equation \eqref{eq:51}. 
Using the expansion indicated in Eq. \eqref{eq:43},  one
gets from the last term in Eq. \eqref{eq:51} 
the Eq. \eqref{eq31} in the Appendix.
%

\begin{figure}[t]
  \includegraphics[scale=0.8]{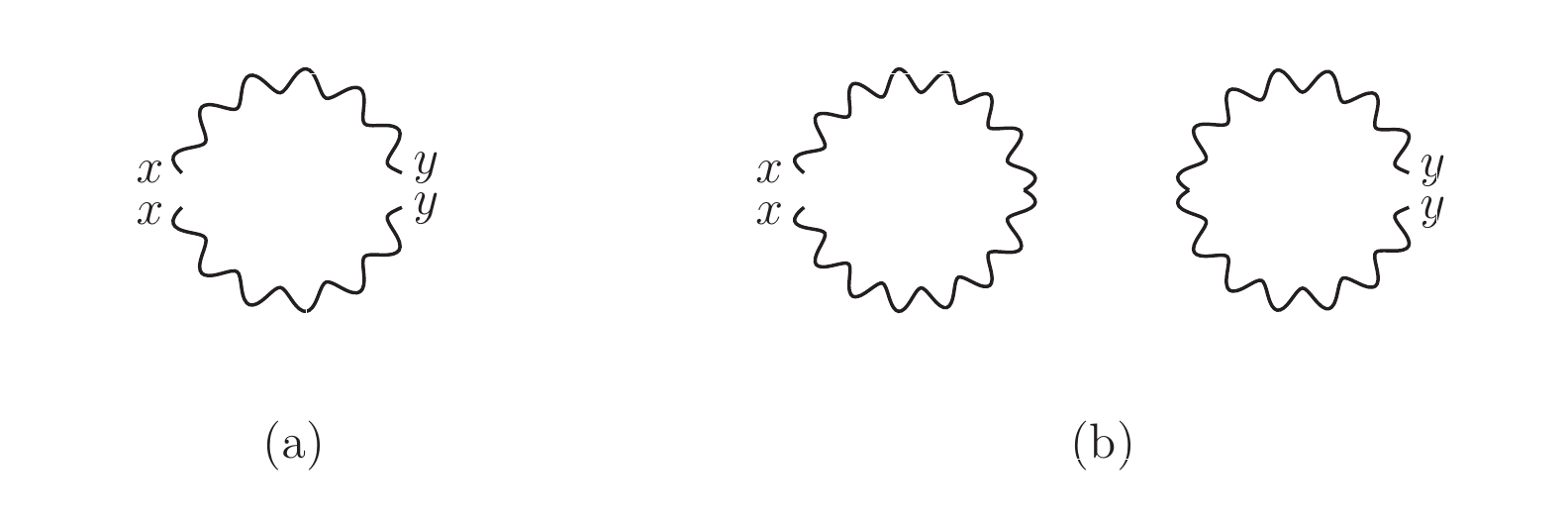} 
\caption{Pinched contributions associated with the last term in
Eq. \eqref{eq:51}.}
\label{fig4}
\end{figure}

We note here that  these composite fields contributions are pinched 
at the spacetime points $x$ and $y$.
The Feynman diagrams  associated with such 
Green's functions are shown in Fig. \ref{fig4}.
The divergent contributions coming from Fig.
(\ref{fig4}a) (there is an additional
graph with $x\leftrightarrow y$ on the left side), 
turn out to add up to a result which agrees 
with that given in Fig. (\ref{fig3}b).
We have also verified this identity 
at one loop order for any dimension $d$ in a
general gauge (see the Appendix).
On the other hand, the contributions coming 
from Fig. (\ref{fig4}b) vanish upon using 
dimensional regularization in momentum space.
Verifying this result beyond order $\kappa^2$ 
becomes exceedingly difficult, as it would involve going beyond one loop order.


We remark that the structural identity \eqref{eq:51}  
resembles the identity \eqref{eq:214}
which holds in the 
Yang-Mills theory.
Therefore, as we have seen in the previous examples,  
the structural identities in the diagonal representation 
of the first order Yang-Mills and gravity theories 
exhibit many similar features, though they are not identical.


\section{Discussion}

We have examined the structural identities which ensure the
self-consistency of the first order formulation of quantum
gravity. Since calculations in this theory are quite involved even at
one loop order, 
we have studied first the structural identities in the diagonal
representation of Yang-Mills theory, which are simpler. It turns out
that these identities in Yang-Mills theory
have many features similar   
to the ones which occur in the diagonal representation of the first order quantum gravity. With this insight, we have compared the forms of the generating functionals $Z[J,  j]$ of
Green's functions in quantum gravity, before and after integrating out the auxiliary field $H_{\mu \nu}^{\lambda}$.
Differentiations of these two forms with respect to $J^{\lambda}_{\mu \nu} $
and $j_{\mu \nu} $ yield a set structural identities 
given in Eqs. \eqref{eq:41} and \eqref{eq:51}
which are complementary
but distinct from the usual Ward identities. These identities show that the Green's functions containing only external external graviton (gluon) fields are the same in the first and second-order formulations.

These identities also lead to  connections between the Green's
functions involving the field $H_{\mu \nu}^{\lambda}$ and the Green's
functions in second-order formulation containing a composite graviton
field that corresponds to the classical value of the auxiliary
field. Eq. \eqref{eq:44} provides
a simple interpretation of the auxiliary field $H_{\mu \nu}^{ \lambda}$. 
An interesting feature is that
the implementation of the structural identities requires 
cancellations between UV divergences which appear in one-loop
diagrams, and the short-distance singularities that occur in the tree graphs
which are ``pinched'' at the same spacetime points. This shows that the
singularities arising at the tree level from the composite graviton field are necessary 
for the first order formulation of quantum gravity to be consistent.
These identities have also a practical utility as they allow us 
to compute more efficiently, in the second-order formulation, 
some involved composite field expectation values in terms of those 
containing the local auxiliary field.

Recently \cite{doi:10.1139/cjp-2019-0037, Brandt:2019ymg}, 
we have introduced a Lagrange multiplier field which restricts the
path integral in quantum gravity to the field configurations that 
satisfy the classical equations of motion. It was shown that such a
method has the effect of eliminating all multi-loop corrections beyond
the one-loop order
and doubling of the usual one-loop contributions.
This makes it possible to renormalize the EH action while retaining unitarity.
Such a treatment was employed both in the second-order as well as in the
first order formulations of quantum gravity. In the later case, 
one may also expect to have a corresponding set of 
structural identities which are necessary for the consistency of the theory. 
This is an interesting issue which requires further study.
 
\begin{acknowledgments}
{This study was financed in part by the Coordena\c{c}\~{a}o de Aperfei\c{c}oamento 
de Pessoal de N\'{\i}vel Superior - Brasil (CAPES) - Finance Code 001.
F. T. B. and J. F. thank CNPq (Brazil) for financial
support. S. M.-F. thanks CAPES (Brazil) for financial support.
D. G. C. M. thanks Roger Macleod for an enlightening discussion.
This work comes as an aftermath of an original 
project developed with
the support of FAPESP (Brazil),  grant number 2018/01073-5.}
\end{acknowledgments}


\appendix*

\section{One loop results}
We employ the same Feynman rules, procedures and conventions 
as in section 3 of \cite{Brandt:2016eaj}, with the replacements 
$\phi^{\mu\nu} \rightarrow \kappa \phi^{\mu\nu}$, 
$G^{\lambda}_{\mu\nu} \rightarrow  \kappa H^{\lambda}_{\mu\nu} $ 
and 
$S \rightarrow S/\kappa^2$ ($S$ is the action), 
so that the coupling constant $\kappa$  is shown explicitly 
in the vertices and in the resulting Green's functions.

\subsection{The general approach for 
the calculation of massless one-loop self-energies}\label{genSE}
Let us consider some generic field theory for fields $\phi_a$,
where $a$ represents a collection of Lorentz indices, or indices for
internal degrees of freedom such as in the case of Yang-Mills theories.
The most general form of the momentum space massless self-energy is
\be\label{genSE1}
\Pi_{a b}(k) = \int \fracd{d^d p}{(2\pi)^d} I_{ab}(p,q)
= \sum_{i=1}^{n} C^i T^i_{ab}(k);  \;\; (q\equiv k+p),
\ee
where $n$ is the number of independent tensors which can be
obtained from the general symmetry properties of $\Pi_{ab}(k)$
(for instance, in the case of the photon self-energy there are the two
independent tensors $\eta_{\mu\nu}$ and $k_\mu k_\nu$).
Upon contracting Eq. \eqref{genSE1} with each of the $n$ tensors, we
obtain $n$ linear equations for the coefficients $C^i$, containing
several scalar integrals of the following type
(using Einstein summation convention for the labels $a$ and $b$)
\be
 \int \fracd{d^d p}{(2\pi)^d} I_{ab}(p,q) T^i_{ab}(k).
\ee
Next, we simplify the $n$ scalars $I_{ab}(p,q) T^i_{ab}(k)$, using the relations
\be
p\cdot k = \frac 1 2 (q^2-p^2-k^2), \;\;
q\cdot k = \frac 1 2 (q^2-p^2+k^2)\;\; \mbox{and} \;\;
p\cdot q = \frac 1 2 (p^2+q^2-k^2)
\ee
so that all the scalar integrals acquire the form
\be
I^{rs} = \int \fracd{d^d p}{(2\pi)^d} 
\fracd{1}{(p^2)^r (q^2)^s} .
\ee
In the simplest cases $r=s=1$. When considering gauge theories with a
general gauge fixing parameter, we can have $r=1,2$ and $s=1,2$.
Since we are using a dimensional regularization procedure, the only
non-vanishing integrals are the following
\begin{subequations}\label{eqI} 
\be\label{eqIa} 
I^{11} = i\fracd{ (k^2)^{d/2-2}}{2^d \pi^{d/2}}
\frac{\Gamma\left(2-\fracd d 2\right) \Gamma\left(\fracd d 2-1\right)^2}
{\Gamma\left(d - 2\right)} \equiv I, 
\ee 
\be \label{eqIb} 
I^{12} = I^{21} = \frac{3-d}{k^2} I,
\ee
\be \label{eqIc} 
I^{22} =  \frac{(3-d)(6-d)}{k^4} I 
\ee
\end{subequations}
(we have a factor of $i$ relative to Eq. (3.31a) of 
\cite{Brandt:2016eaj} which takes into account 
that we are Wick rotating back to Minkowski space).

Once we have all the relevant scalar integrals in Eq. \eqref{eqI}, 
we may  solve the linear system of algebraic equations for constants $C^i$
in  \eqref{genSE1}. In general
this procedure would be of no practical use unless we make use
of computer algebra algorithms, as we have done in the present work
(for example the
tensor basis for the self-energy of the $H$ field has 22 rank 6 tensors).
Using this procedure, we have previously obtained the 
expression for the graviton self-energy in the 
diagonalized first order formalism \cite{Brandt:2016eaj}.

For $d=4-2\epsilon$ Eq. \eqref{eqI} yields the following UV 
pole part
\be\label{CUV} 
I^{UV}  \equiv \fracd{i}{16\pi^2\epsilon}.
\ee 

It is worth mentioning that the present approach is as an example of the Passarino-Veltman reduction method \cite{Passarino:1978jh}.

\subsection{The $H \phi$ self-energy}
Figures (\ref{fig1}a) and (\ref{fig1}b), without the external free 
propagators, are the two contributions for the 
mixed $H\phi$-fields self-energy. Since the internal $H$ field 
propagator in Fig (\ref{fig1}a) has no momentum dependence, 
the loop momentum integration vanishes when using dimensional 
regularization. 
The self-energy contribution from Fig. (\ref{fig1}b) 
can be expressed as 
\be\label{selfHp}
(\Pi^{H\phi})^{\lambda}_{\mu\nu}{}^{\pi\tau} =
\sum_{i=1}^{12} C^{H\phi}_{(i)}
(T_{i}^{H\phi})^{\lambda}_{\mu\nu}{}^{\pi\tau} , 
\ee 
where the tensors 
$(T_{i}^{H\phi})^{\lambda}_{\mu\nu}{}^{\pi\tau}$, 
$i=1 \dots 12$, are 
given by 
\begin{subequations}\label{TensHp}
\begin{eqnarray}
(T^{H\phi}_{1})^{\lambda}_{\mu\nu}{}^{\pi\tau}  & = &
 \fracd{1}{4} \left(k^{\pi} \delta^{\lambda}_{\nu} \delta_{\mu}^{\tau}+k^{\pi} 
\delta^{\lambda}_{\mu}
   \delta_{\nu}^{\tau}+\delta^{\pi}_{ \nu} k^{\tau }
                                                      \delta^{\lambda}_{\mu}
+\delta^{\pi}_{ \mu } k^{\tau} \delta^{\lambda}_{\nu}\right), 
\end{eqnarray}\begin{eqnarray}
(T^{H\phi}_2)^{\lambda}_{\mu\nu}{}^{\pi\tau}  & = &
 \fracd{1}{2}  k^{\lambda } \left(\delta^{\pi}_{ \nu }
                                                    \delta_{\mu}^{\tau}+\delta^{\pi}_{
                                                    \mu } 
\delta_{\nu}^{ \tau }\right), \end{eqnarray}\begin{eqnarray}
(T^{H\phi}_3)^{\lambda}_{\mu\nu}{}^{\pi\tau}  & = & 
k^{\lambda }  \eta^{\pi \tau } \eta_{\mu\nu}, \end{eqnarray}\begin{eqnarray}
(T^{H\phi}_{4})^{\lambda}_{\mu\nu}{}^{\pi\tau}  & = &
 \fracd{1}{2}  \eta_{\mu\nu} \left(k^{\pi} \eta^{\lambda\tau}+\eta^{\lambda\pi} k^{\tau 
   }\right), 
\end{eqnarray}\begin{eqnarray}
(T^{H\phi}_5)^{\lambda}_{\mu\nu}{}^{\pi\tau}  & = &
 \fracd{1}{4}  \left(\delta^{\pi}_{ \nu } k_{\mu}
                                                    \eta^{\lambda\tau}+\eta^{\lambda\pi} 
k_{\mu}
   \delta_{\nu}^{\tau}+\delta^{\pi}_{ \mu } k_{\nu}
                                                    \eta^{\lambda\tau}+\eta^{\lambda\pi} 
k_{\nu} \delta_{\mu}^{\tau }\right), 
\end{eqnarray}\begin{eqnarray}
(T^{H\phi}_{6})^{\lambda}_{\mu\nu}{}^{\pi\tau}  & = &
 \fracd{1}{2}  \eta^{\pi \tau } \left(k_{\mu}
                                                      \delta^{\lambda}_{\nu}+k_{\nu} 
\delta^{\lambda}_{ \mu }\right), 
\end{eqnarray}\begin{eqnarray}
(T^{H\phi}_7)^{\lambda}_{\mu\nu}{}^{\pi\tau}  & = &
 \fracd{1}{2 k^2} k^{\pi} k^{\tau} \left(k_{\mu}
                                                    \delta^{\lambda}_{\nu}+k_{\nu} 
\delta^{\lambda}_{ \mu }\right),
\end{eqnarray}\begin{eqnarray}
(T^{H\phi}_8)^{\lambda}_{\mu\nu}{}^{\pi\tau}  & = & 
\fracd{1}{4 k^2} k^{\lambda } \left(k^{\pi} k_{\mu} \delta_{\nu}^{\tau}+k^{\pi} k_{\nu}
   \delta_{\mu}^{\tau}+\delta^{\pi}_{ \nu } k_{\mu} k^{\tau }+\delta^{\pi}_{ \mu
                                                  } k_{\nu} k^{\tau}\right), 
 \end{eqnarray}\begin{eqnarray}  
(T^{H\phi}_9)^{\lambda}_{\mu\nu}{}^{\pi\tau}  & = &
 \fracd{1}{2 k^2} k_{\mu} k_{\nu} \left(k^{\pi} \eta^{\lambda\tau}+\eta^{\lambda\pi} 
   k^{\tau }\right), 
\end{eqnarray}\begin{eqnarray}
(T^{H\phi}_{10})^{\lambda}_{\mu\nu}{}^{\pi\tau}  & = &
\fracd{1}{k^4}
k^{\lambda }   k_{\mu} k_{\nu}  k^{\pi}   k^{\tau}, 
\end{eqnarray}\begin{eqnarray}
(T^{H\phi}_{11})^{\lambda}_{\mu\nu}{}^{\pi\tau}  & = &
 \fracd{1}{k^2}  k^{\lambda } \eta_{\mu\nu}  k^{\pi} k^{\tau}, 
\end{eqnarray}\begin{eqnarray}
(T^{H\phi}_{12})^{\lambda}_{\mu\nu}{}^{\pi\tau}  & = & 
\fracd{1}{k^2} k^{\lambda } k_{\mu} k_{\nu} \eta^{\pi \tau }.
\end{eqnarray}
\end{subequations}

Using the Feynman rules given in Ref. \cite{Brandt:2016eaj}, 
we obtain the equivalent of $I_{ab}(p,q)$ in Eq. \eqref{genSE1}. 
Next, using the general approach described in subsection (\ref{genSE}) 
we obtain the 
coefficients for the $H\phi$ self-energy shown in Table (\ref{tabHp}). 
These expressions  have an UV part, which arises when $d=4-2 \epsilon$ 
and $\epsilon\rightarrow 0$, given by the numbers in Table (\ref{tabHpUV})

\begin{table}[hbt!]
{\footnotesize
\begin{tabular}{|c||c|c|c|}
\hline 
 & $1$  &  $(\xi-1)$    &  $(\xi-1)^2$ \\ 
\hline 
\hline 
 $C^{H\phi}_{(1)} $ & $ -\frac{1}{16 (d-1)} $ & $ -\frac{1}{16 (d-1)} $ & ${ 0} $ \\ & & & \\ $
 C^{H\phi}_{(2)} $ & $ -\frac{(d-2) (d+2)}{16 (d-1)} $ & $ \frac{3 d^2-18 d+16}{16 (d-1)} $ & $ -\frac{d^3-8 d^2+22 d-14}{16 (d-1)} $ \\ & & & \\ $
 C^{H\phi}_{(3)} $ & $ \frac{d^2+2 d+2}{32 (d-2) (d-1)} $ & $ -\frac{7 d^2-41 d+32}{32 (d-2) (d-1)} $ & $ \frac{5 d^3-37 d^2+96 d-60}{64 (d-2) (d-1)} $ \\ & & & \\ $
 C^{H\phi}_{(4)} $ & ${ 0} $ & $ -\frac{1}{32 (d-1)} $ & ${ 0} $ \\ & & & \\ $
 C^{H\phi}_{(5)} $ & $ \frac{d^2+d-1}{16 (d-1)} $ & $ \frac{5-d}{8} $ & $ \frac{d-2}{8 (d-1)} $ \\ & & & \\ $
 C^{H\phi}_{(6)} $ & $ -\frac{d^2+6 d-4}{32 (d-2) (d-1)} $ & $ \frac{d-5}{8 (d-2)} $ & $ \frac{d-6}{32 (d-1)} $ \\ & & & \\ $
 C^{H\phi}_{(7)} $ & $ \frac{(d-2)^2}{32 (d-1)} $ & $ -\frac{d-2}{16 (d-1)} $ & $ \frac{(d-2)^2}{32 (d-1)} $ \\ & & & \\ $
 C^{H\phi}_{(8)} $ & $ \frac{(d-2) (d+2)}{16 (d-1)} $ & $ -\frac{4 d^2-23 d+18}{16 (d-1)} $ & $ \frac{d^3-8 d^2+21 d-12}{8 (d-1)} $ \\ & & & \\ $
 C^{H\phi}_{(9)} $ & $ -\frac{1}{16} (d+2) $ & $ \frac{4 d^2-23 d+22}{32 (d-1)} $ & $ -\frac{d-2}{8 (d-1)} $ \\ & & & \\ $
 C^{H\phi}_{(10)} $ & ${ 0} $ & $ \frac{(d-4) d}{16 (d-1)} $ & $ -\frac{(d-4)^2 d}{32 (d-1)} $ \\ & & & \\ $
 C^{H\phi}_{(11)} $ & $ -\frac{d^2-10 d+4}{32 (d-1)} $ & $ -\frac{3 \left(d^2-6 d+4\right)}{32 (d-1)} $ & $ \frac{1}{64} (d-2)^2 $ \\ & & & \\ $
 C^{H\phi}_{(12)} $ & $ \frac{d}{8 (d-2) (d-1)} $ & $ \frac{d-2}{32 (d-1)} $ & $ -\frac{(d-4)^2 d}{32 (d-2) (d-1)}$ \\ & & & \\
\hline
\end{tabular}}
\caption{Coefficients for the mixed $H\phi$ self energy 
(see Eq. \eqref{selfHp}) 
in units of $i \kappa^2 k^2I$, 
where $I$ is given by Eq. \eqref{eqIa}.}
\label{tabHp}
\end{table}

\begin{table}[hbt!]
{\footnotesize
\begin{tabular}{ccc}
\begin{tabular}{|c||c|c|c|}
\hline 
 & $1$ & $(\xi-1)$   & $(\xi-1)^2$ \\  
\hline 
\hline  
$C^{H\phi}_{(1)} $ & $ -\frac{1}{48} $ & $ -\frac{1}{48} $ & ${ 0} $ \\ & & & \\ $
 C^{H\phi}_{(2)} $ & $ -\frac{1}{4} $ & $ -\frac{1}{6} $ & $ -\frac{5}{24} $ \\ & & & \\ $
 C^{H\phi}_{(3)} $ & $ \frac{13}{96} $ & $ \frac{5}{48} $ & $ \frac{13}{96} $ \\ & & & \\ $
 C^{H\phi}_{(4)} $ & ${ 0} $ & $ -\frac{1}{96} $ & ${ 0} $ \\ & & & \\ $
 C^{H\phi}_{(5)} $ & $ \frac{19}{48} $ & $ \frac{1}{8} $ & $ \frac{1}{12} $ \\ & & & \\ $
 C^{H\phi}_{(6)} $ & $ -\frac{3}{16} $ & $ -\frac{1}{16} $ & $ -\frac{1}{48} $ 
\\ & & & \\ 
\hline
\end{tabular}
& &
\begin{tabular}{|c||c|c|c|}
\hline 
 & $1$ & $(\xi-1)$   & $(\xi-1)^2$ \\  
\hline 
\hline  
$C^{H\phi}_{(7)} $ & $ \frac{1}{24} $ & $ -\frac{1}{24} $ & $ \frac{1}{24} $ \\ & & & \\ $
 C^{H\phi}_{(8)} $ & $ \frac{1}{4} $ & $ \frac{5}{24} $ & $ \frac{1}{3} $ \\ & & & \\ $
 C^{H\phi}_{(9)} $ & $ -\frac{3}{8} $ & $ -\frac{1}{16} $ & $ -\frac{1}{12} $ \\ & & & \\ $
 C^{H\phi}_{(10)} $ & ${ 0} $ & ${ 0} $ & ${ 0} $ \\ & & & \\ $
 C^{H\phi}_{(11)} $ & $ \frac{5}{24} $ & $ \frac{1}{8} $ & $ \frac{1}{16} $ \\ & & & \\ $
 C^{H\phi}_{(12)} $ & $ \frac{1}{12} $ & $ \frac{1}{48} $ & ${ 0} $ 
\\ & & & \\ 
\hline
\end{tabular}
\end{tabular}}
\caption{The UV parts of the coefficients for the mixed $H\phi$ self energy
(see Eq. \eqref{selfHp}) 
in units of $i \kappa^2 k^2I^{UV}$, 
where $I^{UV}$ is given by Eq. \eqref{eqIa}.}
\label{tabHpUV}
\end{table}

\subsection{The $H$-field self-energy}
Figures (\ref{fig3}a) and (\ref{fig3}b), without the external free 
propagators, are the two contributions for the 
$H$-field self-energy. Since the internal $H$ field 
propagator in Fig (\ref{fig3}a) has no momentum dependence, 
the loop momentum integration vanishes when using dimensional 
regularization. 
The self-energy contribution from Fig. (\ref{fig3}b) 
can be expressed as 
\be\label{selfHH}
(\Pi^{HH})^{\lambda}_{\mu\nu}{}^\rho_{\pi\tau} =
\sum_{i=1}^{22} C^{HH}_{(i)}
(T_{i}^{HH})^{\lambda}_{\mu\nu}{}^\rho_{\pi\tau} , 
\ee 
where the tensors 
$(T_{i}^{HH})^{\lambda}_{\mu\nu}{}^\rho_{\pi\tau}$, 
$i=1 \dots 22$, are 
given by 
\begin{subequations}\label{tensHH}
\begin{eqnarray}
 (T^{HH}_{1}) ^{\lambda}_{\mu\nu}{}^\rho_{\pi\tau}
  & = & \fracd{1}{4}  \left(\delta_\pi^{\rho } \delta^{\lambda}_{\nu} \eta_{\mu\tau}+\delta_\pi^{\rho }
   \delta^{\lambda}_{\mu} \eta_{\nu\tau}+\eta_{\pi \nu } \delta^{\lambda}_{\mu} \delta^\rho_{ \tau }+\eta_{\pi \mu 
   } \delta^{\lambda}_{\nu} \delta^\rho_{ \tau }\right), 
\end{eqnarray}\begin{eqnarray} 
 (T^{HH}_{2})^{\lambda}_{\mu\nu}{}^\rho_{\pi\tau}  & = & \fracd{1}{2}  \eta^{\lambda \rho } \left(\eta_{\pi\nu } \eta_{\mu\tau}+\eta_{\pi \mu } \eta_{\nu \tau }\right), 
\end{eqnarray}\begin{eqnarray} 
 (T^{HH}_{3})^{\lambda}_{\mu\nu}{}^\rho_{\pi\tau}  & = &  \eta_{\pi \tau 
                                                       } \eta^{\lambda \rho }
                                                       \eta_{\mu\nu}, 
\end{eqnarray}\begin{eqnarray} 
 (T^{HH}_{4})^{\lambda}_{\mu\nu}{}^\rho_{\pi\tau}  & = & \fracd{1}{4}  \left(\eta_{\pi\nu } \delta^{\lambda}_{\tau} \delta_\mu^{\rho }+\delta_{\pi}^{ \lambda }
   \delta_\mu^{\rho } \eta_{\nu \tau}+\eta_{\pi \mu } \delta^{\lambda}_{\tau} \delta_\nu^{\rho }+\delta_{\pi}^{ \lambda } \eta_{\mu\tau} \delta_\nu^{\rho }\right), 
\end{eqnarray}\begin{eqnarray} 
 (T^{HH}_{5})^{\lambda}_{\mu\nu}{}^\rho_{\pi\tau}  & = & \fracd{1}{4}  \left(\delta_\pi^{\rho } \delta^{\lambda}_{\tau} \eta_{\mu\nu}+\delta_{\pi}^{ \lambda }
   \eta_{\mu\nu} \delta^\rho_{ \tau }+\eta_{\pi \tau } \delta^{\lambda}_{\nu} \delta_\mu^{\rho }+\eta_{\pi \tau }
   \delta^{\lambda}_{\mu} \delta_\nu^{\rho }\right), 
\end{eqnarray}\begin{eqnarray} 
 (T^{HH}_{6})^{\lambda}_{\mu\nu}{}^\rho_{\pi\tau}  & = & \fracd{1}{4 k^4}  \left(k_{\pi } k^{\lambda } k_{\mu} k_{\nu } \delta^\rho_{ \tau }+\delta_\pi^{\rho 
   } k^{\lambda } k_{\mu} k_{\nu } k_{\tau }+k_{\pi } k_{\mu} k^{\rho } k_{\tau }
   \delta^{\lambda}_{\nu}+k_{\pi } k_{\nu } k^{\rho } k_{\tau } \delta^{\lambda}_{\mu}\right), 
\end{eqnarray}\begin{eqnarray} 
(T^{HH}_{7})^{\lambda}_{\mu\nu}{}^\rho_{\pi\tau}  & = & \fracd{1}{4 k^4}  k^{\lambda } k^{\rho } \left(k_{\pi } k_{\mu} \eta_{\nu \tau}+k_{\pi }
   k_{\nu } \eta_{\mu\tau}+\eta_{\pi  \nu } k_{\mu} k_{\tau }+\eta_{\pi \mu } k_{\nu } k_{\tau 
   }\right), 
\end{eqnarray}\begin{eqnarray} 
 (T^{HH}_{8})^{\lambda}_{\mu\nu}{}^\rho_{\pi\tau}  & = & \fracd{1}{4 k^4} 
                                                       \left(k_{\pi }
                                                       k_{\mu} k_{\nu 
                                                       } k^{\rho }
                                                       \delta^{\lambda}_{
                                                       \tau 
                                                       }+\delta_{\pi}^{ \lambda } k_{\mu} k_{\nu } k^{\rho } k_{\tau }+k_{\pi } k^{\lambda } k_{\mu} k_{\tau 
   } \delta_\nu^{\rho }+k_{\pi } k^{\lambda } k_{\nu } k_{\tau } \delta_\mu^{\rho }\right), 
\end{eqnarray}\begin{eqnarray} 
 (T^{HH}_{9})^{\lambda}_{\mu\nu}{}^\rho_{\pi\tau}  & = & \fracd{1}{2 k^4}  k^{\lambda } k^{\rho } \left(\eta_{\pi \tau } k_{\mu} k_{\nu }+k_{\pi }
   k_{\tau } \eta_{\mu\nu}\right), 
\end{eqnarray}\begin{eqnarray} 
 (T^{HH}_{10})^{\lambda}_{\mu\nu}{}^\rho_{\pi\tau}  & = & \fracd{1}{4 k^2}  \left(k_{\pi } \delta^\rho_{ \tau }+\delta_\pi^{\rho } k_{\tau }\right) \left(k_{\mu 
   } \delta^{\lambda}_{\nu}+k_{\nu } \delta^{\lambda}_{\mu}\right), 
\end{eqnarray}\begin{eqnarray} 
 (T^{HH}_{11})^{\lambda}_{\mu\nu}{}^\rho_{\pi\tau}  & = & \fracd{1}{4 k^2}  \eta^{\lambda \rho } \left(k_{\pi } k_{\mu} \eta_{\nu \tau}+k_{\pi } k_{\nu }
   \eta_{\mu\tau}+\eta_{\pi  \nu } k_{\mu} k_{\tau }+\eta_{\pi \mu } k_{\nu } k_{\tau }\right), 
\end{eqnarray}\begin{eqnarray} 
 (T^{HH}_{12})^{\lambda}_{\mu\nu}{}^\rho_{\pi\tau}  & = & \fracd{1}{4 k^2}  \left(k_{\pi } \delta^{\lambda}_{\tau}+\delta_{\pi}^{ \lambda } k_{\tau }\right) 
   \left(k_{\mu} \delta_\nu^{\rho }+k_{\nu } \delta_\mu^{\rho }\right), 
\end{eqnarray}\begin{eqnarray} 
 (T^{HH}_{13})^{\lambda}_{\mu\nu}{}^\rho_{\pi\tau}  & = & \fracd{1}{4 k^2}  \left(\delta_\pi^{\rho } k_{\mu} k_{\nu } \delta^{\lambda}_{\tau}+\delta_{\pi}^{ \lambda }
   k_{\mu} k_{\nu } \delta^\rho_{ \tau }+k_{\pi } k_{\tau }
                                                          \delta^{\lambda}_{\nu} 
\delta_\mu^{\rho }+k_{\pi } k_{\tau } \delta^{\lambda}_{\mu} \delta_\nu^{\rho }\right), 
\end{eqnarray}\begin{eqnarray} 
 (T^{HH}_{14})^{\lambda}_{\mu\nu}{}^\rho_{\pi\tau}  & = & \fracd{1}{2 k^2}  \eta^{\lambda \rho } \left(\eta_{\pi \tau } k_{\mu} k_{\nu }+k_{\pi } k_{\tau 
   } \eta_{\mu\nu}\right), 
\end{eqnarray}\begin{eqnarray} 
 (T^{HH}_{15})^{\lambda}_{\mu\nu}{}^\rho_{\pi\tau}  & = & \fracd{1}{8 k^2}  \left(\delta_\pi^{\rho } k^{\lambda } k_{\mu} \eta_{\nu \tau}+k_{\pi } k^{\rho 
   } \delta^{\lambda}_{\mu} \eta_{\nu \tau}+\delta_\pi^{\rho }
                                                          k^{\lambda }
                                                          k_{\nu } 
\eta_{\mu \tau 
   }+\eta_{\pi \nu } k^{\lambda } k_{\mu} \delta^\rho_{ \tau }\right. 
\\ && \\
&+& \left. \eta_{\pi \mu } k^{\lambda } k_{\nu }
   \delta^\rho_{ \tau }+k_{\pi } k^{\rho } \delta^{\lambda}_{\nu}
    \eta_{\mu\tau}+
\eta_{\pi \nu } k^{\rho }
   k_{\tau } \delta^{\lambda}_{\mu}+\eta_{\pi \mu } k^{\rho } k_{\tau } \delta^{\lambda}_{\nu}\right), 
\end{eqnarray}\begin{eqnarray} 
 (T^{HH}_{16})^{\lambda}_{\mu\nu}{}^\rho_{\pi\tau}  & = & \fracd{1}{2 k^2}  k^{\lambda } k^{\rho } \left(\eta_{\pi \nu } \eta_{\mu\tau}+\eta_{\pi \mu }
   \eta_{\nu \tau}\right), 
\end{eqnarray}\begin{eqnarray} 
 (T^{HH}_{17})^{\lambda}_{\mu\nu}{}^\rho_{\pi\tau}  & = & \fracd{1}{4 k^2}  \left(\eta_{\pi \tau } k^{\lambda } k_{\mu} \delta_\nu^{\rho }+\eta_{\pi \tau }
   k^{\lambda } k_{\nu } \delta_\mu^{\rho }+k_{\pi } k^{\rho } \delta^{\lambda}_{\tau} \eta_{\mu \nu 
   }+\delta_{\pi}^{ \lambda } k^{\rho } k_{\tau } \eta_{\mu\nu}\right), 
\end{eqnarray}\begin{eqnarray} 
 (T^{HH}_{18})^{\lambda}_{\mu\nu}{}^\rho_{\pi\tau}  & = & \fracd{1}{8 k^2}  \left(k_{\pi } k^{\lambda } \eta_{\mu\tau} \delta_\nu^{\rho }+\eta_{\pi \mu }
   k^{\lambda } k_{\tau } \delta_\nu^{\rho }+k_{\pi } k^{\lambda } \delta_\mu^{\rho } \eta_{\nu \tau 
   }+\eta_{\pi \nu } k_{\mu} k^{\rho } \delta^{\lambda}_{\tau}
\right. 
\\ && \\ 
&+& \left.\delta_{\pi}^{ \lambda } k_{\mu} k^{\rho }
   \eta_{\nu \tau}+\eta_{\pi \mu } k_{\nu } k^{\rho } \delta^{\lambda}_{\tau}+\delta_{\pi}^{\lambda } k_{\nu 
   } k^{\rho } \eta_{\mu\tau}+\eta_{\pi \nu } k^{\lambda } k_{\tau } \delta_\mu^{\rho }\right), 
\end{eqnarray}\begin{eqnarray} 
 (T^{HH}_{19})^{\lambda}_{\mu\nu}{}^\rho_{\pi\tau}  & = & \fracd{1}{4 k^2}  \left(k_{\pi } k^{\lambda } \eta_{\mu\nu} \delta^\rho_{ \tau }+\eta_{\pi \tau }
   k_{\mu} k^{\rho } \delta^{\lambda}_{\nu}+\eta_{\pi \tau } k_{\nu }
                                                          k^{\rho } 
\delta^{\lambda}_{ \mu 
   }+\delta_\pi^{\rho } k^{\lambda } k_{\tau } \eta_{\mu\nu}\right), 
\end{eqnarray}\begin{eqnarray} 
(T^{HH}_{20})^{\lambda}_{\mu\nu}{}^\rho_{\pi\tau}  & = & 
\fracd{1}{k^6} k_{\pi } k^{\lambda } k_{\mu} k_{\nu } k^{\rho }
                                                      k_{\tau }, 
\end{eqnarray}\begin{eqnarray} 
 (T^{HH}_{21})^{\lambda}_{\mu\nu}{}^\rho_{\pi\tau}  & = & \fracd{1}{k^4}
                                                       k_{\pi } k_{\mu 
                                                       } k_{\nu }
                                                       k_{\tau }
                                                       \eta^{\lambda 
                                                       \rho }, 
\end{eqnarray}\begin{eqnarray} 
 (T^{HH}_{22})^{\lambda}_{\mu\nu}{}^\rho_{\pi\tau}  & = &
                                                       \fracd{1}{k^2}\eta_{\pi \tau } k^{\lambda } k^{\rho } \eta_{\mu\nu} . 
\end{eqnarray} 
\end{subequations}

Using the Feynman rules given in Ref. \cite{Brandt:2016eaj}, 
we obtain the equivalent of $I_{ab}(p,q)$ in Eq. \eqref{genSE1}. 
Next, using the general approach described in subsection (\ref{genSE}) 
we obtain the 
coefficients for the $HH$ self-energy shown in Table (\ref{tableHH}). 
These expressions  have an UV part, which arises when $d=4-2 \epsilon$
and $\epsilon\rightarrow 0$, given by the numbers in Table (\ref{tabHHUV}) 

\begin{table}[hbt!]
{\footnotesize
\begin{tabular}{cc}
\begin{tabular}{|c||c|c|c|}
\hline 
 & $1$ & $(\xi-1)$   & $(\xi-1)^2$ \\  
\hline 
\hline 
$C^{HH}_{(1)} $ & $ -\frac{1}{2 (d-2)} $ & $ \frac{3 (d-3) d}{16 (d-2) (d-1)} $ & $ \
-\frac{1}{4 (d-1) (d+1)} $ \\ & & & \\
 $C^{HH}_{(2)} $ & $ -\frac{d^2-2 d-2}{4 (d-2) (d-1)} $ & $ -\frac{4 d^3-17 d^2+31 \
d-32}{16 (d-2) (d-1)} $ & $ -\frac{d+2}{8 (d-1) (d+1)} $ \\ & & & \\
 $C^{HH}_{(3)} $ & $ \frac{1}{2 (d-1)} $ & $ \frac{d (3 d-5)}{16 (d-2) (d-1)} $ & $ \
\frac{d}{16 (d-1) (d+1)} $ \\ & & & \\
 $C^{HH}_{(4)} $ & $ \frac{d^2-2}{4 (d-2) (d-1)} $ & $ -\frac{4 d^3-23 d^2+33 \
d-16}{16 (d-2) (d-1)} $ & $ -\frac{1}{4 (d-1) (d+1)} $ \\ & & & \\
 $C^{HH}_{(5)} $ & $ -\frac{1}{(d-2) (d-1)} $ & $ \frac{d^2-3 d-8}{8 (d-2) (d-1)} $ & $ \
\frac{d}{4 (d-1) (d+1)} $ \\ & & & \\
 $C^{HH}_{(6)} $ & ${ 0} $ & $ -\frac{d-4}{2 (d-1)} $ & $ \frac{(d-4) (d-2) d}{4 (d-1) \
(d+1)} $ \\ & & & \\
 $C^{HH}_{(7)} $ & ${ 0} $ & $ -\frac{(d-4) \left(d^2-31 d+24\right)}{16 (d-1)} $ & $ \
-\frac{(d-4) \left(4 d^3-16 d^2-3 d+14\right)}{4 (d-1) (d+1)} $ \\ & & & \\
 $C^{HH}_{(8)} $ & ${ 0} $ & $ \frac{(d-4)^2 (d-3)}{8 (d-1)} $ & $ -\frac{(d-4) (d-2) \
(d+2)}{4 (d-1) (d+1)} $ \\ & & & \\
 $C^{HH}_{(9)} $ & ${ 0} $ & $ 4-d $ & $ \frac{(d-4) \left(4 d^3-17 d^2-4 d+20\right)}{8 \
(d-1) (d+1)} $ \\ & & & \\
 $C^{HH}_{(10)} $ & $ \frac{d^2-4 d+2}{2 (d-2) (d-1)} $ & $ \frac{3 d^2-12 d+8}{4 \
(d-2) (d-1)} $ & $ \frac{(d-2) d}{4 (d-1) (d+1)} $ \\ & & & \\
$C^{HH}_{(11)} $ & $ \frac{d \left(d^2-d-4\right)}{4 (d-2) (d-1)} $ & $ \frac{d^3+15 \
d^2-50 d+16}{16 (d-2) (d-1)} $ & $ \frac{(d-2) (2 d+1)}{4 (d-1)
                                 (d+1)}$   \\ & & & \\
\hline 
\end{tabular}
& 
\begin{tabular}{|c||c|c|c|}
\hline 
 & $1$ & $(\xi-1)$   & $(\xi-1)^2$ \\  
\hline 
\hline  
 $C^{HH}_{(12)} $ & $ \frac{d^3-2 d^2-4 d+4}{4 (d-2) (d-1)} $ & $ -\frac{(d-4) \
\left(d^2-4 d+2\right)}{4 (d-2) (d-1)} $ & $ \frac{(d-2) d}{4 (d-1) (d+1)} $ \\ & & & \\
 $C^{HH}_{(13)} $ & $ -\frac{(d-4) d}{2 (d-2) (d-1)} $ & $ \frac{3 d^3-23 d^2+50 \
d-16}{8 (d-2) (d-1)} $ & $ -\frac{(d-2) (d+2)}{4 (d-1) (d+1)} $ \\ & & & \\
 $C^{HH}_{(14)} $ & $ -\frac{d}{2 (d-1)} $ & $ \frac{d^2-11 d+12}{8 (d-1)} $ & $ \
-\frac{(d-2) (d+2)}{8 (d-1) (d+1)} $ \\ & & & \\
 $C^{HH}_{(15)} $ & $ \frac{d^2}{2 (d-2) (d-1)} $ & $ -\frac{3 d^3-15 d^2+18 d-16}{8 \
(d-2) (d-1)} $ & $ -\frac{d-2}{2 (d-1) (d+1)} $ \\ & & & \\
 $C^{HH}_{(16)} $ & $ \frac{2 d^2-d-2}{4 (d-1)} $ & $ -\frac{11 d^2-89 d+72}{16 \
(d-1)} $ & $ \frac{4 d^4-28 d^3+55 d^2+32 d-52}{8 (d-1) (d+1)} $ \\ & & & \\
 $C^{HH}_{(17)} $ & $ \frac{d}{2 (d-1)} $ & $ -\frac{(d-7) d}{8 (d-1)} $ & $ \frac{(d-2) \
d}{4 (d-1) (d+1)} $ \\ & & & \\
 $C^{HH}_{(18)} $ & $ -\frac{2 d^2+d-2}{2 (d-1)} $ & $ \frac{11 d^2-73 d+72}{8 (d-1)} \
$ & $ -\frac{(d-2) (2 d+3)}{2 (d-1) (d+1)} $ \\ & & & \\
 $C^{HH}_{(19)} $ & $ -1 $ & $ -1 $ & $ -\frac{(d-2) (d+2)}{4 (d-1) (d+1)} $ \\ & & & \\
 $C^{HH}_{(20)} $ & ${ 0} $ & ${ 0} $ & $ \frac{(d-6) (d-4) (d-2) d}{16 (d-1) (d+1)} $ \\ & & & \\
 $C^{HH}_{(21)} $ & ${ 0} $ & $ -\frac{(d-4) (d-3) d}{16 (d-1)} $ & $ \frac{(d-4) (d-2) \
d}{16 (d-1) (d+1)} $ \\ & & & \\
 $C^{HH}_{(22)} $ & ${ 1}$  & $ \frac{3 d-4}{4 (d-1)} $ & $ \frac{8 d^3-23 d^2-10 d+24}{16 \
(d-1) (d+1)} $ \\ & & & \\
\hline
\end{tabular}
\end{tabular}}
\caption{Coefficients for the $H$-field self energy 
(see Eq. \eqref{selfHH}) 
in units of $\kappa^2 k^2I$, 
where $I$ is given by Eq. \eqref{eqIa}.}
\label{tableHH}
\end{table}

\begin{table}[hbt!]
{\footnotesize
\begin{tabular}{ccc}
\begin{tabular}{|c||c|c|c|}
\hline 
 & $1$ & $(\xi-1)$   & $(\xi-1)^2$ \\  
\hline 
\hline 
 $C^{HH}_{(1)} $ & $ -\frac{1}{4} $ & $ \frac{1}{8} $ & $ -\frac{1}{60} $ \\ &&& \\
 $C^{HH}_{(2)} $ & $ -\frac{1}{4} $ & $ -\frac{19}{24} $ & $ -\frac{1}{20} $ \\ &&& \\
 $C^{HH}_{(3)} $ & $ \frac{1}{6} $ & $ \frac{7}{24} $ & $ \frac{1}{60} $ \\ &&& \\
 $C^{HH}_{(4)} $ & $ \frac{7}{12} $ & $ -\frac{1}{24} $ & $ -\frac{1}{60} $ \\ &&& \\
 $C^{HH}_{(5)} $ & $ -\frac{1}{6} $ & $ -\frac{1}{12} $ & $ \frac{1}{15} $ \\ &&& \\
 $C^{HH}_{(6)} $ & ${ 0} $ & ${ 0} $ & ${ 0} $ \\ &&& \\
 $C^{HH}_{(7)} $ & ${ 0} $ & ${ 0} $ & ${ 0} $ \\ &&& \\
 $C^{HH}_{(8)} $ & ${ 0} $ & ${ 0} $ & ${ 0} $ \\ &&& \\
 $C^{HH}_{(9)} $ & ${ 0} $ & ${ 0} $ & ${ 0} $ \\ &&& \\
 $C^{HH}_{(10)} $ & $ \frac{1}{6} $ & $ \frac{1}{3} $ & $ \frac{2}{15} $ \\ &&& \\
 $C^{HH}_{(11)} $ & $ \frac{4}{3} $ & $ \frac{5}{4} $ & $ \frac{3}{10} $ \\ &&& \\
\hline 
\end{tabular}
& &
\begin{tabular}{|c||c|c|c|}
\hline 
 & $1$ & $(\xi-1)$   & $(\xi-1)^2$ \\  
\hline 
\hline 
$C^{HH}_{(12)} $ & $ \frac{5}{6} $ & ${ 0} $ & $ \frac{2}{15} $ \\ &&& \\
 $C^{HH}_{(13)} $ & ${ 0} $ & $ \frac{1}{6} $ & $ -\frac{1}{5} $ \\ &&& \\
 $C^{HH}_{(14)} $ & $ -\frac{2}{3} $ & $ -\frac{2}{3} $ & $ -\frac{1}{10} $ \\ &&& \\
 $C^{HH}_{(15)} $ & $ \frac{4}{3} $ & $ -\frac{1}{6} $ & $ -\frac{1}{15} $ \\ &&& \\
 $C^{HH}_{(16)} $ & $ \frac{13}{6} $ & $ \frac{9}{4} $ & $ \frac{47}{30} $ \\ &&& \\
 $C^{HH}_{(17)} $ & $ \frac{2}{3} $ & $ \frac{1}{2} $ & $ \frac{2}{15} $ \\ &&& \\
 $C^{HH}_{(18)} $ & $ -\frac{17}{3} $ & $ -\frac{11}{6} $ & $ -\frac{11}{15} $ \\ &&& \\
 $C^{HH}_{(19)} $ & $ -1 $ & $ -1 $ & $ -\frac{1}{5} $ \\ &&& \\
 $C^{HH}_{(20)} $ & ${ 0} $ & ${ 0} $ & ${ 0} $ \\ &&& \\
 $C^{HH}_{(21)} $ & ${ 0} $ & ${ 0} $ & ${ 0} $ \\ &&& \\
 $C^{HH}_{(22)} $ & ${ 1}$  & $ \frac{2}{3} $ & $ \frac{8}{15} $ \\ &&& \\
\hline
\end{tabular}
\end{tabular}}
\caption{Coefficients for the UV part of the 
  $H$-field self energy (see Eq. \eqref{selfHH}) 
in units of $\kappa^2 k^2 I^{UV}$,  where $I^{UV}$ is given by Eq. \eqref{CUV}.}
\label{tabHHUV}
\end{table}

\subsection{Propagators}

\subsubsection{Mixed $H\phi$ propagator}

The mixed $H\phi$ propagator 
$(M^{-1} \Pi^{H\phi} {\cal D})^{\lambda}_{\mu\nu}{}^{\pi\tau}$
can also be expressed in terms of the tensor basis  
in Eq. \eqref{TensHp} as 
\be\label{propHp1}
(M^{-1} \Pi^{H\phi} {\cal D})^{\lambda}_{\mu\nu}{}^{\pi\tau} =
\sum_{i=1}^{12} P^{H\phi}_{(i)} (T_{i}^{H\phi})^{\lambda}_{\mu\nu}{}^{\pi\tau}.
\ee 
The coefficients $P^{H\phi}_{(i)}$ are obtained 
by solving the system of 12 algebraic equations which results from 
the contractions of Eq. \eqref{propHp1} with 
$(T_{j}^{H\phi})^{\lambda}_{\mu\nu}{}^{\pi\tau}$, $j=1\dots 12$. 
A straightforward computer algebra calculation generates relations 
between $ P^{H\phi}_{(i)}$ and  $ C^{H\phi}_{(i)} $. Then, using the results 
for $C^{H\phi}_{(i)}$ given in Table (\ref{tabHp})  we obtain the entries of Table 
(\ref{tablePropHphi}) for the mixed $H\phi$ propagator. 
Table (\ref{tablePropHphiUV}) shows the UV part 
of the mixed $H\phi$ propagator, which arises when $d=4-2 \epsilon$ 
and $\epsilon\rightarrow 0$, obtained from 
Table (\ref{tablePropHphi}) making $d=4$.

\begin{table}[hbt!]
{\footnotesize
\begin{tabular}{|c||c|c|c|}
\hline 
 & $1$ & $(\xi-1)$   & $(\xi-1)^2$ \\  
\hline 
\hline 
 $P^{H\phi}_{(1)} $ & ${ 0} $ & $ -\frac{i}{16 (d-1)} $ & $ -\frac{i}{16 \
(d-1)} $ \\ & & & \\ $ 
 P^{H\phi}_{(2)} $ & $ \frac{i \left(2 d^2+d-5\right)}{16 \
(d-1)} $ & $ -\frac{i (5 (d-6) d+26)}{16 (d-1)} \
$ & $ \frac{i (d ((d-8) d+24)-18)}{16 (d-1)} $ \\ & & & \\ $ 
 P^{H\phi}_{(3)} $ & $ -\frac{i (d ((d-5) d+19)-6)}{32 \
(d-2) (d-1)} $ & $ \frac{i \left(23 d^2-147 \
d+122\right)}{64 \left(d^2-3 d+2\right)} $ & $ -\frac{i \
(d (d (10 d-73)+193)-134)}{64 (d-2) \
(d-1)} $ \\ & & & \\ $ 
 P^{H\phi}_{(4)} $ & $ -\frac{i}{16 (d-1)} $ & $ -\frac{3 i}{32 (d-1)} \
$ & $ -\frac{i}{32 (d-1)} $ \\ & & & \\ $ 
 P^{H\phi}_{(5)} $ & $ -\frac{i \left(d^2-4\right)}{8 (d-1)} $ & $ \
\frac{i (3 (d-6) d+16)}{8 (d-1)} $ & $ -\frac{i \
(d ((d-8) d+22)-14)}{8 (d-1)} $ \\ & & & \\ $ 
 P^{H\phi}_{(6)} $ & $ \frac{i ((d-6) d-3)}{16 (d-1)} $ & $ \
\frac{1}{32} i {{(4 d -27)}} $ & $ -\frac{i (d ((d-7) \
d+21)-17)}{32 (d-1)} $ \\ & & & \\ $ 
 P^{H\phi}_{(7)} $ & $ -\frac{i ((d-10) d+4)}{16 (d-1)} $ & $ \
-\frac{i (d (3 d-26)+26)}{16 (d-1)} $ & $ \frac{i \
(d ((d-9) d+34)-28)}{32 (d-1)} $ \\ & & & \\ $ 
 P^{H\phi}_{(8)} $ & $  -\frac{1}{8} { {i (d+2)}} $ & $ \frac{i (d (4 \
d-23)+24)}{16 (d-1)} $ & $ -\frac{i (3 d-10)}{16 \
(d-1)} $ \\ & & & \\ $ 
 P^{H\phi}_{(9)} $ & $ \frac{i (d+2) (2 d-3)}{16 (d-1)} $ & $ \
-\frac{i (d (12 d-71)+54)}{32 (d-1)} $ & $ \frac{i \
(d (4 (d-8) d+85)-50)}{32 (d-1)} $ \\ & & & \\ $ 
 P^{H\phi}_{(10)} $ & $ { 0} $ & $ \frac{i (d-6) d}{16 (d-1)} $ & $ \
-\frac{i ((d-5) (d-4) d+4)}{32 (d-1)} $ \\ & & & \\ $ 
 P^{H\phi}_{(11)} $ & $ \frac{i (d ((d-5) d+28)-20)}{32 \
(d-2) (d-1)} $ & $ -\frac{i \left(7 d^2-48 \
d+44\right)}{16 \left(d^2-3 d+2\right)} $ & $ \frac{i \
(d (d (5 d-39)+110)-80)}{32 (d-2) \
(d-1)} $ \\ & & & \\ $ 
 P^{H\phi}_{(12)} $ & $ -\frac{i (d-2)}{32 (d-1)} $ & $ \frac{i \
((d-9) d+2)}{64 (d-1)} $ & $ -\frac{i (d \
(d (2 d-15)+43)-34)}{64 (d-1)} $ \\ & & & \\  
\hline
\end{tabular}}
\caption{Coefficients for the mixed $H \phi$ propagator
(see Eq. \eqref{propHp1}) in units of $\kappa^2 I$, 
where $I$ is given by Eq. \eqref{eqIa}.}
\label{tablePropHphi}
\end{table}

\begin{table}[hbt!]
{\footnotesize
\begin{tabular}{|c||c|c|c|}
\hline 
 & $1$ & $(\xi-1)$   & $(\xi-1)^2$ \\  
\hline 
\hline 
$ P^{H\phi}_{(1)} $ & $ { 0} $ & $ -\frac{i}{48} $ & $ -\frac{i}{48} $ \\ $
 P^{H\phi}_{(2)} $ & $ \frac{31 i}{48} $ & $ \frac{7 i}{24} $ & $ \frac{7 i}{24} $ \\ $
 P^{H\phi}_{(3)} $ & $ -\frac{9 i}{32} $ & $ -\frac{49 i}{192} $ & $ -\frac{55 i}{192} $ \\ $
 P^{H\phi}_{(4)} $ & $ -\frac{i}{48} $ & $ -\frac{i}{32} $ & $ -\frac{i}{96} $ \\ $
 P^{H\phi}_{(5)} $ & $ -\frac{i}{2} $ & $ -\frac{i}{3} $ & $ -\frac{5 i}{12} $ \\ $
 P^{H\phi}_{(6)} $ & $ -\frac{11 i}{48} $ & $ -\frac{11 i}{32} $ & $ -\frac{19 i}{96} $ \\ $
 P^{H\phi}_{(7)} $ & $ \frac{5 i}{12} $ & $ \frac{5 i}{8} $ & $ \frac{7 i}{24} $ \\ $
 P^{H\phi}_{(8)} $ & $ -\frac{3 i}{4} $ & $ -\frac{i}{12} $ & $ -\frac{i}{24} $ \\ $
 P^{H\phi}_{(9)} $ & $ \frac{5 i}{8} $ & $ \frac{19 i}{48} $ & $ \frac{17 i}{48} $ \\ $
 P^{H\phi}_{(10)} $ & $ { 0} $ & $ -\frac{i}{6} $ & $ -\frac{i}{24} $ \\ $
 P^{H\phi}_{(11)} $ & $ \frac{19 i}{48} $ & $ \frac{3 i}{8} $ & $ \frac{7 i}{24} $ \\ $
 P^{H\phi}_{(12)} $ & $ -\frac{i}{48} $ & $ -\frac{3 i}{32} $ & $ -\frac{13 i}{96} $\\
\hline
\end{tabular}}
\caption{UV part of the coefficients for the mixed $H \phi$ propagator
(see Eq. \eqref{propHp1}) 
in units of $\kappa^2 I^{UV}$, 
where $I^{UV}$ is given by Eq. \eqref{CUV}.}
\label{tablePropHphiUV}
\end{table}

\subsubsection{$H$ field  propagator}
The $HH$ propagator 
$(M^{-1} \Pi^{HH} M^{-1})^{\lambda}_{\mu\nu}{}^\rho_{\pi\tau}$
can also be expressed in terms of the tensor basis  
in Eq. \eqref{tensHH} as 
\be\label{propHH1}
(M^{-1} \Pi^{HH} M^{-1})^{\lambda}_{\mu\nu}{}^\rho_{\pi\tau} =
\sum_{i=1}^{22} P^{HH}_{(i)} (T_{i}^{HH})^{\lambda}_{\mu\nu}{}^\rho_{\pi\tau}.
\ee 
The coefficients $P^{HH}_{(i)}$ are obtained 
by solving the system of 22 algebraic equations which results from 
the contractions of Eq. \eqref{propHH1} with 
$(T_{j}^{HH})^{\lambda}_{\mu\nu}{}^\rho_{\pi\tau}$, $j=1\dots 22$. 
A straightforward computer algebra calculation generates the relations 
between $ P^{HH}_{(i)}$ and  $ C^{HH}_{(i)} $. Then, using the results 
for $C^{HH}_{(i)}$ given in Table (\ref{tableHH})  we obtain the entries of Table 
(\ref{tablePropHH}) for the $H$-field propagator. 
Table (\ref{tablePropUVHH}) shows the UV part 
of the $H$-field propagator, which arises when $d=4-2 \epsilon$ 
and $\epsilon\rightarrow 0$, obtained from 
Table (\ref{tablePropHH}) making $d=4$.

\begin{table}[hbt!]
{\footnotesize
\begin{tabular}{|c||c|c|c|}
\hline 
 & $1$ & $(\xi-1)$   & $(\xi-1)^2$ \\  
\hline 
\hline 
$ P^{HH}_{(1)} $ & $ \frac{1}{2 (d-1)} $ & $ \frac{d (3 d-5)}{16 (d-2) (d-1)} $ & $ \
\frac{d}{16 (d-1) (d+1)} $ \\ & & & \\ $
 P^{HH}_{(2)} $ & $ -\frac{2 d^2-3 d-4}{8 (d-2) (d-1)} $ & $ -\frac{2 d^3-7 d^2+15 \
d-20}{16 (d-2) (d-1)} $ & $ -\frac{3 d+4}{32 (d-1) (d+1)} $ \\ & & & \\ $
 P^{HH}_{(3)} $ & $ \frac{1}{8 (d-2) (d-1)} $ & $ \frac{d^2-2 d+4}{16 (d-2) (d-1)} $ & $ \
-\frac{3 d+4}{64 (d-1) (d+1)} $ \\ & & & \\ $
 P^{HH}_{(4)} $ & $ \frac{d+1}{4 (d-1)} $ & $ \frac{3 d^2-d-8}{16 (d-2) (d-1)} $ & $ \
\frac{d}{16 (d-1) (d+1)} $ \\ & & & \\ $
 P^{HH}_{(5)} $ & $ -\frac{1}{2 (d-2)} $ & $ -\frac{d^2-d+4}{8 (d-2) (d-1)} $ & $ \
\frac{d}{16 (d-1) (d+1)} $ \\ & & & \\ $
 P^{HH}_{(6)} $ & ${ 0} $ & $ \frac{4-d}{2} $ & $ \frac{(d-4) \left(4 d^3-17 d^2-4 \
d+20\right)}{16 (d-1) (d+1)} $ \\ & & & \\ $
 P^{HH}_{(7)} $ & ${ 0} $ & $ -\frac{(d-4) (d-3) d}{16 (d-1)} $ & $ \frac{(d-4) (d-2) \
d}{16 (d-1) (d+1)} $ \\ & & & \\ $
 P^{HH}_{(8)} $ & ${ 0} $ & $ \frac{(d-4) (d-3) (d-2)}{8 (d-1)} $ & $ -\frac{(d-4) (d-2) \
(3 d+4)}{16 (d-1) (d+1)} $ \\ & & & \\ $
 P^{HH}_{(9)} $ & ${ 0} $ & $ \frac{(d-4) (d+2)}{8 (d-2) (d-1)} $ & $ -\frac{(d-4) \left(5 \
d^3-14 d^2-20 d+8\right)}{32 (d-2) (d-1) (d+1)} $ \\ & & & \\ $
 P^{HH}_{(10)} $ & ${ 1}$  & $ \frac{3 d-4}{4 (d-1)} $ & $ \frac{8 d^3-23 d^2-10 d+24}{16 \
(d-1) (d+1)} $ \\ & & & \\ $
 P^{HH}_{(11)} $ & $ \frac{d^2}{4 (d-1)} $ & $ -\frac{3 d^2-41 d+32}{16 (d-1)} $ & $ \
\frac{4 d^4-28 d^3+57 d^2+30 d-56}{16 (d-1) (d+1)} $ \\ & & & \\ $
 P^{HH}_{(12)} $ & $ \frac{(d-2) (d+1)}{4 (d-1)} $ & $ \frac{5-d}{2} $ & $ \frac{4 \
d^4-28 d^3+53 d^2+34 d-48}{16 (d-1) (d+1)} $ \\ & & & \\ $
 P^{HH}_{(13)} $ & $ \frac{d}{2 (d-1)} $ & $ -\frac{d^2-9 d+6}{8 (d-1)} $ & $ \
\frac{(d-2) (3 d+2)}{16 (d-1) (d+1)} $ \\ & & & \\ $
 P^{HH}_{(14)} $ & ${ 0} $ & $ \frac{1-d}{8} $ & $ -\frac{(d-2) (d+2)}{32 (d-1) (d+1)} $ \\ & & & \\ $
 P^{HH}_{(15)} $ & $ -\frac{d}{2 (d-1)} $ & $ \frac{d^2-11 d+12}{8 (d-1)} $ & $ \
-\frac{(d-2) (d+2)}{8 (d-1) (d+1)} $ \\ & & & \\ $
 P^{HH}_{(16)} $ & $ \frac{4 d^3-7 d^2-8 d+8}{8 (d-2) (d-1)} $ & $ -\frac{9 d^3-87 \
d^2+202 d-120}{16 (d-2) (d-1)} $ & $ \frac{4 d^4-28 d^3+69 d^2+18 d-80}{32 (d-1) \
(d+1)} $ \\ & & & \\ $
 P^{HH}_{(17)} $ & $ \frac{d}{2 (d-2) (d-1)} $ & $ \frac{d^3-5 d^2+8 d+4}{8 (d-2) \
(d-1)} $ & $ -\frac{(d-2) (d+2)}{16 (d-1) (d+1)} $ \\ & & & \\ $
 P^{HH}_{(18)} $ & $ -d-1 $ & $ \frac{11 d^2-81 d+72}{8 (d-1)} $ & $ -\frac{4 d^4-28 \
d^3+59 d^2+30 d-64}{8 (d-1) (d+1)} $ \\ & & & \\ $
 P^{HH}_{(19)} $ & $ \frac{3}{d-2} $ & $ -\frac{7 d^2-44 d+36}{4 (d-2) (d-1)} $ & $ \
\frac{3 \left(4 d^4-25 d^3+44 d^2+28 d-48\right)}{16 (d-2) (d-1) (d+1)} $ \\ & & & \\ $
 P^{HH}_{(20)} $ & ${ 0} $ & ${ 0} $ & $ \frac{(d-6) (d-4) (d-2) d}{64 (d-1) (d+1)} $ \\ & & & \\ $
 P^{HH}_{(21)} $ & ${ 0} $ & $ -\frac{(d-4) \left(d^2-12 d+12\right)}{16 (d-1)} $ & $ \
-\frac{(d-4) \left(16 d^3-69 d^2-10 d+72\right)}{64 (d-1) (d+1)} $ \\ & & & \\ $
 P^{HH}_{(22)} $ & 
$
-\frac{d \left(2 d^2+d-24\right)+20}{8 (d-2)^2 (d-1)}
$ &
$\frac{d (d (5 d-49)+130)-88}{8 (d-2)^2 (d-1)}$
& 
$\frac{d (d (d ((97-8 d) d-390)+468)+360)-576}{64 (d-2)^2 \left(d^2-1\right)}$
\\ &&& \\
\hline 
\end{tabular}}
\caption{Coefficients for the $H$-field propagator 
(see Eq. \eqref{propHH1}) 
in units of $\kappa^2 k^2I$, 
where $I$ is given by Eq. \eqref{eqIa}.}
\label{tablePropHH}
\end{table}

\begin{table}[hbt!]
{\footnotesize
\begin{tabular}{ccc}
\begin{tabular}{|c||c|c|c|}
\hline 
 & $1$ & $(\xi-1)$   & $(\xi-1)^2$ \\  
\hline 
\hline  
$P^{HH}_{(1)} $ & $ \frac{1}{6} $ & $ \frac{7}{24} $ & $ \frac{1}{60} $ \\ &&& \\ $
 P^{HH}_{(2)} $ & $ -\frac{1}{3} $ & $ -\frac{7}{12} $ & $ -\frac{1}{30} $ \\ &&& \\ $
 P^{HH}_{(3)} $ & $ \frac{1}{48} $ & $ \frac{1}{8} $ & $ -\frac{1}{60} $ \\ &&& \\ $
 P^{HH}_{(4)} $ & $ \frac{5}{12} $ & $ \frac{3}{8} $ & $ \frac{1}{60} $ \\ &&& \\ $
 P^{HH}_{(5)} $ & $ -\frac{1}{4} $ & $ -\frac{1}{3} $ & $ \frac{1}{60} $ \\ &&& \\ $
 P^{HH}_{(6)} $ & ${ 0} $ & ${ 0} $ & ${ 0} $ \\ &&& \\ $
 P^{HH}_{(7)} $ & ${ 0} $ & ${ 0} $ & ${ 0} $ \\ &&& \\ $
 P^{HH}_{(8)} $ & ${ 0} $ & ${ 0} $ & ${ 0} $ \\ &&& \\ $
 P^{HH}_{(9)} $ & ${ 0} $ & ${ 0} $ & ${ 0} $ \\ &&& \\ $
 P^{HH}_{(10)} $ & ${ 1}$  & $ \frac{2}{3} $ & $ \frac{8}{15} $ \\ &&& \\ $
 P^{HH}_{(11)} $ & $ \frac{4}{3} $ & $ \frac{7}{4} $ & $ \frac{13}{15}
                                                       $ \\ &&& \\
\hline
\end{tabular}
& &
\begin{tabular}{|c||c|c|c|}
\hline 
 & $1$ & $(\xi-1)$   & $(\xi-1)^2$ \\  
\hline 
\hline  
$ P^{HH}_{(12)} $ & $ \frac{5}{6} $ & $ \frac{1}{2} $ & $ \frac{7}{10} $ \\ &&& \\ $
 P^{HH}_{(13)} $ & $ \frac{2}{3} $ & $ \frac{7}{12} $ & $ \frac{7}{60} $ \\ &&& \\ $
 P^{HH}_{(14)} $ & ${ 0} $ & $ -\frac{3}{8} $ & $ -\frac{1}{40} $ \\ &&& \\ $
 P^{HH}_{(15)} $ & $ -\frac{2}{3} $ & $ -\frac{2}{3} $ & $ -\frac{1}{10} $ \\ &&& \\ $
 P^{HH}_{(16)} $ & $ \frac{5}{2} $ & $ \frac{4}{3} $ & $ \frac{41}{60} $ \\ &&& \\ $
 P^{HH}_{(17)} $ & $ \frac{1}{3} $ & $ \frac{5}{12} $ & $ -\frac{1}{20} $ \\ &&& \\ $
 P^{HH}_{(18)} $ & $ { -5} $ & $ -\frac{19}{6} $ & $ -\frac{29}{15} $ \\ &&& \\ $
 P^{HH}_{(19)} $ & $ \frac{3}{2} $ & $ \frac{7}{6} $ & $ \frac{6}{5} $ \\ &&& \\ $
 P^{HH}_{(20)} $ & ${ 0} $ & ${ 0} $ & ${ 0} $ \\ &&& \\ $
 P^{HH}_{(21)} $ & ${ 0} $ & ${ 0} $ & ${ 0} $ \\ &&& \\ $
 P^{HH}_{(22)} $ & $ -\frac{17}{24} $ & $ -\frac{1}{3} $ & $
                                                           \frac{1}{120}
                                                           $ \\  &&& \\
\hline 
\end{tabular}
\end{tabular}}
\caption{The UV pole part of the coefficients for the $H$-field
propagator (see Eq. \eqref{propHH1}) 
in units of $\kappa^2 k^2I^{UV}$, 
where $I^{UV}$ is given by Eq. \eqref{CUV}.}
\label{tablePropUVHH}
\end{table}

\subsection{Explicit verification of the structural identities}

The right side of Eq.~\eqref{eq:41}, at order $ \kappa^{2} $, can be written as 
\begin{equation}\label{eq:41a}
\kappa 
{\cal M}{}^{\lambda}_{\mu\nu}{}^{\rho}_{\alpha\beta}{}_{\gamma\delta}  
\langle 
0| T \phi^{\gamma\delta}(x) \phi^{\alpha\beta}_{,\rho}(x) 
\phi^{\pi \tau}_{}(y) | 0 
\rangle,
\end{equation} 
where ${\cal M}$ is defined in such a way that
\be\label{A29a}
-\kappa (M^{-1}(\eta) M(\phi) M^{-1}(\eta))
^{\lambda}_{\mu\nu}{}^\rho_{\alpha\beta} \equiv
\kappa \phi^{\gamma\delta} 
{\cal M}{}^{\lambda}_{\mu\nu}{}^\rho_{\alpha\beta\;\gamma\delta}  
\ee 
with $M(\phi)$ given by \eqref{eq:36BM2016}. 

In momentum space, Eq. \eqref{eq:41a} can be written as
\begin{equation}\label{A16a} 
-i\kappa
{\cal M}{}^{\lambda}_{\mu\nu}{}^{\rho}_{\alpha\beta}{}_{\gamma\delta}  
\left[
\int \fracd{d^d p}{(2\pi)^d} \, p_{\rho}
\mathcal{D}^{\alpha\beta}{}^{ \sigma_1 \theta_1 }(p) 
\mathcal{D}^{\gamma \delta}{}^{\sigma_{2} \theta_{2} } (q) 
\mathcal{V}_{\sigma_{1} \theta_{1}\sigma_{2} \theta_{2}\sigma_{3}
        \theta_{3}}(-p,q,-k)\right]
\mathcal{D}^{\sigma_{3} \theta_{3}}{}^{ \pi \tau}{}(k) .
\end{equation} 
We are using the same notation employed for the self-energies 
($p$ is the integration momentum, $k$ is an external momentum and $q=p+k$); 
${\cal D}^{\mu\nu \rho\sigma}(p)$ is the graviton propagator and $ \mathcal{V}_{\mu \nu \alpha \beta 
\gamma \delta}(p,q,r)$ is the cubic graviton vertex given respectively by the 
Eqs.~(3.25a) and (3.25e) of \cite{Brandt:2016eaj} 
\footnote{Both the three graviton interaction vertex  
and the propagator are the same as 
in the second-order formalism from the expansion of Eq. \eqref{eq:43}.}. 
Since ${\cal M}$ is just a combination of products of $\eta$s and
$\delta$s, each of the several terms in 
Eq. \eqref{A16a} can be cast in the same form as \eqref{propHp1},
in terms of the tensor basis given by Eqs.~\eqref{TensHp}. 
After a straightforward calculation, we have obtained a result which
coincides with the 
one-loop contribution to the 
mixed $H\phi$ propagator
(the same structure constants shown in Table \eqref{tablePropHphi}),
which confirms the identity \eqref{eq:41} for any dimension and gauge parameter.

Similarly, the second term on the right side of Eq. \eqref{eq:51}, at order $\kappa^2$, can be written as 
\be\label{eq31}
\kappa^2 
{\cal M}{}^{\lambda}_{\mu\nu}{}^{\rho_1}_{\pi_1\tau_1}{}_{\gamma_1\delta_1}  
\langle 
0| T \phi^{\gamma_1\delta_1}(x) \phi^{\pi_1\tau_1}_{,\rho_1}(x) 
\phi^{\gamma_2\delta_2}(y) \phi^{\pi_2\tau_2}_{,\rho_2}(y) | 0 
\rangle 
{\cal M}{}^\rho_{\pi\tau}{}^{\rho_2}_{\pi_2\tau_2}{}_{\gamma_2\delta_2}.
\ee   

In momentum space, Eq. \eqref{eq31} can be written as
\be\label{A16} 
-\kappa^2 
{\cal M}{}^{\lambda}_{\mu\nu}{}^{\rho_1}_{\pi_1\tau_1}{}_{\gamma_1\delta_1}  
\left\{
\int \fracd{d^d p}{(2\pi)^d}
\left[
{\cal D}^{\pi_1\tau_1 \gamma_2\delta_2}(p) 
{\cal D}^{\gamma_1\delta_1 \pi_2\tau_2}(q) p_{\rho_1} 
-q_{\rho_1} {\cal D}^{\gamma_1\delta_1 \gamma_2\delta_2}(p) 
{\cal D}^{\pi_1\tau_1 \pi_2\tau_2}(q)  
\right] q_{\rho_2} \right\}
{\cal M}{}^\rho_{\pi\tau}{}^{\rho_2}_{\pi_2\tau_2}{}_{\gamma_2\delta_2}.
\ee 
Eq. \eqref{A16} can also be cast in the same form as \eqref{propHH1},
in terms of the tensor basis given by Eqs. \eqref{tensHH}. 
After a straightforward calculation, we have obtained a result which
coincides with the one-loop contribution to the 
$H$-field propagator
(the same structure constants shown in Table \eqref{tablePropHH}),
which confirms the identity \eqref{eq:51} for any dimension and gauge parameter.

We point out that these structural identities relate elements of the
basic Feynman rules, in each formalism, in a non-trivial way. There is also a practical implication since these identities allow
one to compute some rather involved composite field expectation values in a much more efficient way by using the auxiliary field instead.


\end{document}